\documentclass[journal]{IEEEtran}
\usepackage{cite}
\usepackage{amsmath,amssymb,amsfonts}
\usepackage{booktabs}
\usepackage{mathrsfs}
\usepackage{algorithmic}
\usepackage{graphicx}
\usepackage{epstopdf}
\usepackage{textcomp}
\usepackage{xcolor}
\usepackage{algorithm}
\usepackage{algorithmic}
\usepackage{multicol}
\usepackage{stfloats}
\floatname{algorithm}{Algorithm}

\ifCLASSINFOpdf
\else
\fi
\begin{document}
%
\title{Anypath Routing Protocol Design via Q-Learning for Underwater Sensor Networks}
%
%
%

\author{Yuan~Zhou,~\IEEEmembership{Senior Member,~IEEE,}
        Tao~Cao,
        and~Wei~Xiang,~\IEEEmembership{Senior Member,~IEEE}

\thanks{Yuan Zhou is a visiting fellow with the Department of Electrical Engineering, Princeton University, Princeton, NJ,
USA. (e-mail: yuanzhou@princeton.edu)}

\thanks{Yuan Zhou and Tao Cao are with the School of Electrical and Information Engineering,
Tianjin University, Tianjin 300072, China. (e-mail: zhouyuan@tju.edu.cn, caotao@tju.edu.cn)}

\thanks{Wei~Xiang is with the College of Science, Technology and Engineering, James Cook University, Cairns QLD 4870 Australia. (e-mail: wei.xiang@jcu.edu.cn)}
}

%
%

\markboth{IEEE INTERNET OF THINGS JOURNAL}%
{Shell \MakeLowercase{\textit{et al.}}: Bare Demo of IEEEtran.cls for IEEE Journals}
%



\maketitle

\begin{abstract}
 As a promising technology in the Internet of Underwater Things, underwater sensor networks have drawn a widespread attention from both academia and industry. However, designing a routing protocol for underwater sensor networks is a great challenge due to high energy consumption and large latency in the underwater environment. This paper proposes a Q-learning-based localization-free anypath routing (QLFR) protocol to prolong the lifetime as well as reduce the end-to-end delay for underwater sensor networks. Aiming at optimal routing policies, the Q-value is calculated by jointly considering the residual energy and depth information of sensor nodes throughout the routing process. More specifically, we define two reward functions (i.e., depth-related and energy-related rewards) for Q-learning with the objective of reducing latency and extending network lifetime. In addition, a new holding time mechanism for packet forwarding is designed according to the priority of forwarding candidate nodes. Furthermore, a mathematical analysis is presented to analyze the performance of the proposed routing protocol. Extensive simulation results demonstrate the superiority performance of the proposed routing protocol in terms of the end-to-end delay and the network lifetime.
\end{abstract}


%
\IEEEpeerreviewmaketitle

\section{Introduction}
\begin{figure}[t]
\centerline{\includegraphics[width = 1\columnwidth]{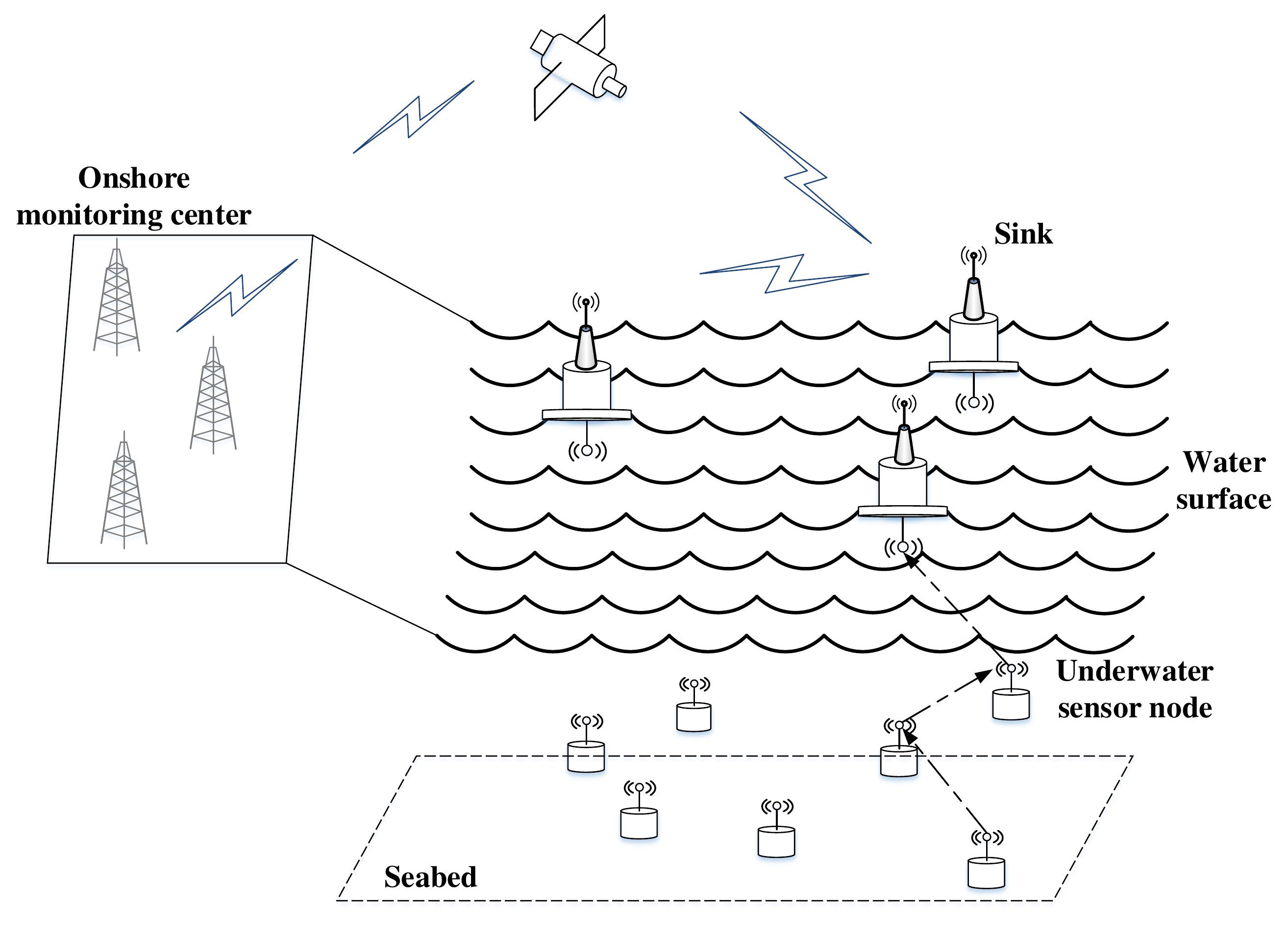}}
\caption{Schematic of the UWSN.}
\label{fig1}
\end{figure}

\IEEEPARstart{T}{he} Internet of Things (IoT), as a promising networking paradigm, can render convenient and efficient services for a wide range of application domains without manual intervention \cite{1,2,3,4}. With an increasing interest in observing and exploring marine resources, the concept of IoT has extended to underwater environments, forming the so-called Internet of Underwater Things (IoUT) \cite{5,6,7}. The IoUT is committed to providing interconnectivity among intelligent underwater devices to monitor vast unexplored underwater areas \cite{8}. As critical infrastructure in the IoUT, underwater sensor networks (UWSN) have found numerous underwater applications \cite{9}, such as offshore oil exploration and extraction, environmental observation for scientific exploration, ocean disaster prevention, mine recognition, and navigation assistance \cite{10,11,12,13,14}. Fig.~\ref{fig1} illustrates the architecture of the UWSN. Due to the harsh underwater environment and high deployment costs, deploying UWSNs is much more challenging than deploying terrestrial wireless sensor networks (WSNs) \cite{15,16,17}.

Acoustic communications are preferred for UWSNs because they provide longer propagation distances \cite{18}. However, the propagation speed of underwater acoustic waves is approximately 1500 m/s \cite{19}, resulting in large propagation latency for underwater networking services. In addition, energy efficiency has also been a major design concern for UWSNs due to the high communication energy cost \cite{20} and the limited energy \cite{21}.

Anypath routing (a.k.a. opportunistic routing)\cite{23} is considered an effective strategy for both energy efficiency and propagation latency in UWSNs. Using anypath routing, a subset of neighboring nodes is selected by the sender as the forwarding candidates according to certain criteria, e.g., energy efficiency and latency. Meanwhile, these selected forwarding candidates are assigned different priorities so that they can cooperate with each other to elect the appropriate next-hop forwarder which minimizes redundant packet transmission.

To address the energy efficiency problem, many anypath routing protocols for UWSNs favor the shortest path to forward sensory data so as to minimize energy consumption \cite{22,23,24}. However, some nodes may be over-burdened with transmitting too many packets and become hot spots in these methods. These hot spots may fail prematurely due to energy exhaustion, disrupting the operations and shortening the lifetime of the entire UWSN \cite{25,26,27}. To extend the network lifetime, it is proposed encourage packet forwarding via the nodes with more residual energy \cite{28,29,30}.

To tackle the latency issue, some anypath routing protocols are proposed by employing greedy approaches to reduce the end-to-end delay \cite{31,32,33}. However, the greedy approaches do not consider the long-term rewards. Thus, using these algorithms, the next hop chosen by the current node may not be the global optimal one for the whole routing path. Moreover, full-dimensional location information is usually required in the routing design process of these methods, which may not be practical in the underwater environment.

In order to simultaneously tackle the issues of high end-to-end latency and low energy efficiency in UWSNs, this paper proposes a novel Q-learning-based localization-free anypath routing (QLFR) protocol. By using the Q-learning algorithm, the QLFR protocol takes the long-term reward into account, and thus is able to make a global optimal routing decision. Two reward functions, i.e., the depth-related and energy-related rewards, are designed for Q-learning. More elaborately, with the depth-related reward function, the proposed routing protocol does not require knowledge of the full-dimensional localization of nodes; instead, it only needs to know the depth information which can be easily obtained with a hydraulic pressure gauge. With the energy-related reward function, the nodes with more residual energy are more likely to forward data packets. As a result, the workload among sensor nodes is more balanced. We also design a new holding time mechanism for anypath packet forwarding. With such a mechanism, the forwarding candidate nodes are scheduled to transmit data packages in accordance with their priority levels. In addition, a multipath suppression scheme is proposed to further reduce unnecessary transmissions and to improve energy efficiency.

The main contributions of this paper are summarized as follows.
\begin{enumerate}
\item[1)] Different from other Q-learning based routing protocols which simply choose the neighbor with the maximum Q-value as the next hop, we design a priority mechanism according to the Q-value when choosing an appropriate next hop;
    \item[2)] We design a new holding time mechanism for anypath routing, according to the priority optimized by Q-learning. To the best of our knowledge, we are the first to use the reinforcement learning technique to design anypath routing protocols in UWSNs; and
        \item[3)] We propose a multipath suppression scheme to further reduce unnecessary transmissions while ensuring a high packet delivery ratio.
\end{enumerate}

The rest of this paper is organized as follow. Section II provides an overview of the related work on routing protocols for UWSNs. In Section III, we first present the network topology architecture, and then model the UWSN routing problem in the general framework of Q-leaning. Section IV describes the proposed QLFR algorithm in detail and Section V elaborate on the corresponding routing protocol. A theoretical analysis of the protocol performance is presented in Section VI. Simulation results and discussions are reported in Section VII. Finally, concluding remarks are drawn in Section VIII. Symbols used in the following sections are listed and defined in Table I.

\begin{table}[t]\label{Table 1}
\centering
\normalsize
\caption{Symbol list}
 \scalebox{0.9}{
 \begin{tabular}{ll}
  \toprule
  Parameters & Definition \\
  \midrule
  $\mathcal{S}, \mathcal{A}, \mathcal{P}, \mathcal{R}$             &         Set of states, actions, transition probabilities\\
  &and rewards in reinforcement learning \\
  &theory, respectively             \\
  $S_t, A_t$            &         State and action in timestep $t$             \\
  $r, \pi$              &          Reward function and policy in \\
  &reinforcement learning theory           \\
  $V_{\pi}(s), V(s)$           &Expected reward when starting in state $s$, \\
  & following policy $\pi$ and the corresponding \\
  &approximated value in Q-learning \\
  $ Q_{\pi}(s,a), Q(s,a) $  &Expected reward after taking action $a$\\
  &in state $s$, following policy $\pi$ and the \\
  &corresponding approximated value in \\
  &Q-learning \\
  $\gamma, \alpha$  &Discount factor and learning rate in \\
  &calculating Q-value \\
  $c_e$  &Residual energy-related reward function \\
  $c_d$  &Depth-related reward function \\
  $e_{\rm res}(s_i)$     &             Residual energy of node $s_i$          \\
  $e_{\rm ini}(s_i)$ & Initial energy of node $s_i$ \\
  $depth(s_i)$ &Depth of node $s_i$ \\
  $d(s_i,s_j)$ &Difference between the depth of node  \\
  &$s_i$ and $s_j$\\
  $n$ &The sequence number of a sensor node \\
  &in the priority list \\
  $\tau$   &   Holding time\\
  $k, b$ &Parameters of holding time\\
  $R$ &Maximal transmission range of a node\\
  $v_0$  &Speed of acoustic waves in water \\
  $t_{\rm max}$  &Maximal propagation delay of one hop \\
  $v$  &Movement speed of nodes in UWSN\\
  $A(l,f)$ & Attenuation of underwater acoustic signal \\
  &with frequency $f$ kHz at transmission \\
  &distance $l$ meters \\
  $\overline{SNR(l,f)}$ & Average signal-to-noise ratio \\
  $p(l,f,M)$ & Packet error rate when the size of \\
  &packet is $M$ bits \\
  $P_{s_i-\rm sink}$ & Delivery probability from $s_i$ to sink \\
  $D_{s_is_j}$ & Distance between the node $s_i$ and $s_j$ \\
  $T_{s_i-\rm sink}$ & End-to-end delay from $s_i$ to sink \\
  $\lambda_{s_i}$ & Outgoing traffic of node $s_i$ \\
  $E_{s_i}$ & Energy consumption of node $s_i$ \\
  $\Gamma_{s_i}$ & Lifetime of node $s_i$ \\
  $\Gamma_{\rm net}$ & Network lifetime \\
\bottomrule
\end{tabular}}
\end{table}

\section{Related Work}
Routing protocols tailored for underwater sensor networks have been developed for over a decade \cite{34,35,36,37}. In this section, we give an overview of relevant underwater routing solutions and highlight their characteristics.

At the beginning, routing protocols in UWSNs are designed based on geographic locations of sensor nodes. Jornet \emph{et al.} \cite{38} proposed a distributed approach©¤--the focused beam routing (FBR) protocol. In this method, power control and location information are both involved in the protocol design to select the appropriate next hop forwarder. The Vector-Based Forwarding (VBF) protocol was present in \cite{23}, in which data packets are forwarded in a virtual pipeline with a predefined radius. The virtual pipeline is specified by the routing vector from the position locations of the source node and the its destination. In order to reduce energy consumption, a self-adaptation algorithm is developed, which weighs the benefit of nodes within pipeline to forward packets and makes the nodes with low benefit discard the packets. AHH-VBF \cite{39} was a successor of VBF, in which the radius of virtual pipeline and the transmission power level are both adaptively changed hop by hop to guarantee the transmission reliability in sparse region and optimize energy efficiency. Coutinho \emph{et al.} proposed GEDAR \cite{35} that employs the greedy forwarding strategy to route data packets. This strategy is executed with the location of the current sender, its neighbors, and the sink node on the water surface, to determine the eligible neighboring nodes to continue forwarding the data packet towards the sink. Although these algorithms show decent performance, they assume to know the full-dimensional localization of the sensor nodes, which is still a challenge in UWSNs \cite{40,41}.

To suit the unique property of UWSNs, many localization-free routing protocols have been emerged in recent years. Yan \emph{et al.} \cite{42} proposed the Depth-Based Routing (DBR) protocol, which is the first underwater routing solution that exploits depth information of sensor node to forward data packets. Moreover, a holding time mechanism is designed to help coordinating the transmission of forwarding candidates.  Wahid and Kim \cite{43} extended DBR protocol to an Energy Efficient Depth-Based Routing (EEDBR) protocol, in which both the depth information and the residual energy of sensor nodes are taken into account to select the next hop node. Lee \emph{et al.} \cite{44} reported the Hydrocast anypath routing protocol which also use depth to advance data packets. In Hydrocast, the priority of next-hop node is determined based on the trade-off between link cost and progress of the packet towards the surface. Coutinho \emph{et al.} \cite{45} proposed an energy balancing opportunistic routing protocol EnOR that take both packet advancement and workload balancing into consideration. VAPR \cite{31} combined depth information and hop count to set up next hop data forwarding direction, building a directional routing path to the closest sink. The next-hop forwarding set is selected according to the current data forwarding direction and next-hop data forwarding direction. Guan \emph{et al.} \cite{46} suggested a distance-vector-Based opportunistic routing (DVOR) that uses the hop counts of sensor node toward the destination to seek the shortest routing path. Based on the hop counts, a holding time mechanism is also designed to schedule the packets forwarding. Both VAPR and DVOR used periodic beacons to dynamically establish the routing path, which will cause significant overheads to UWSNs. In \cite{47}, localization-free anypath routing and duty-cycling were symbiotic designed to achieve reliable data transmission and improve energy efficiency for UWSNs. Coutinho \emph{et al.} \cite{48} combined power control and localization-free anypath routing for UWSNs to simultaneously reduce energy consumption and improve data delivery reliability.

Reinforcement learning has been well exploited for designing routing protocols in WSNs \cite{49,50,51,52}, and more recently for routing solutions in UWSNs \cite{53,54,56}. The Q-learning-based adaptive routing (QELAR) protocol proposed in \cite{53} proved that Q-learning is perform well in UWSNs. QELAR defined the reward function based on the residual energy of each sensor nodes and the energy distribution among neighboring nodes. In this protocol, sensor nodes choose the node with more residual energy as the next hop, so that the network lifetime of the network can be extended. Moreover, to improve the reliability of data transmission, a retransmission mechanism after transmission failures is used in QELAR. Q-learning technique was also used to select the most promising forwarders so as to reduce the end-to-end delay in \cite{54,56}.

In this paper, we design a localization-free anypath routing protocol with the reinforcement learning technique (Q-learning) to reduce the end-to-end delay and extend the network lifetime for UWSNs.

\begin{figure}[t]
\centerline{\includegraphics[width = 1\columnwidth]{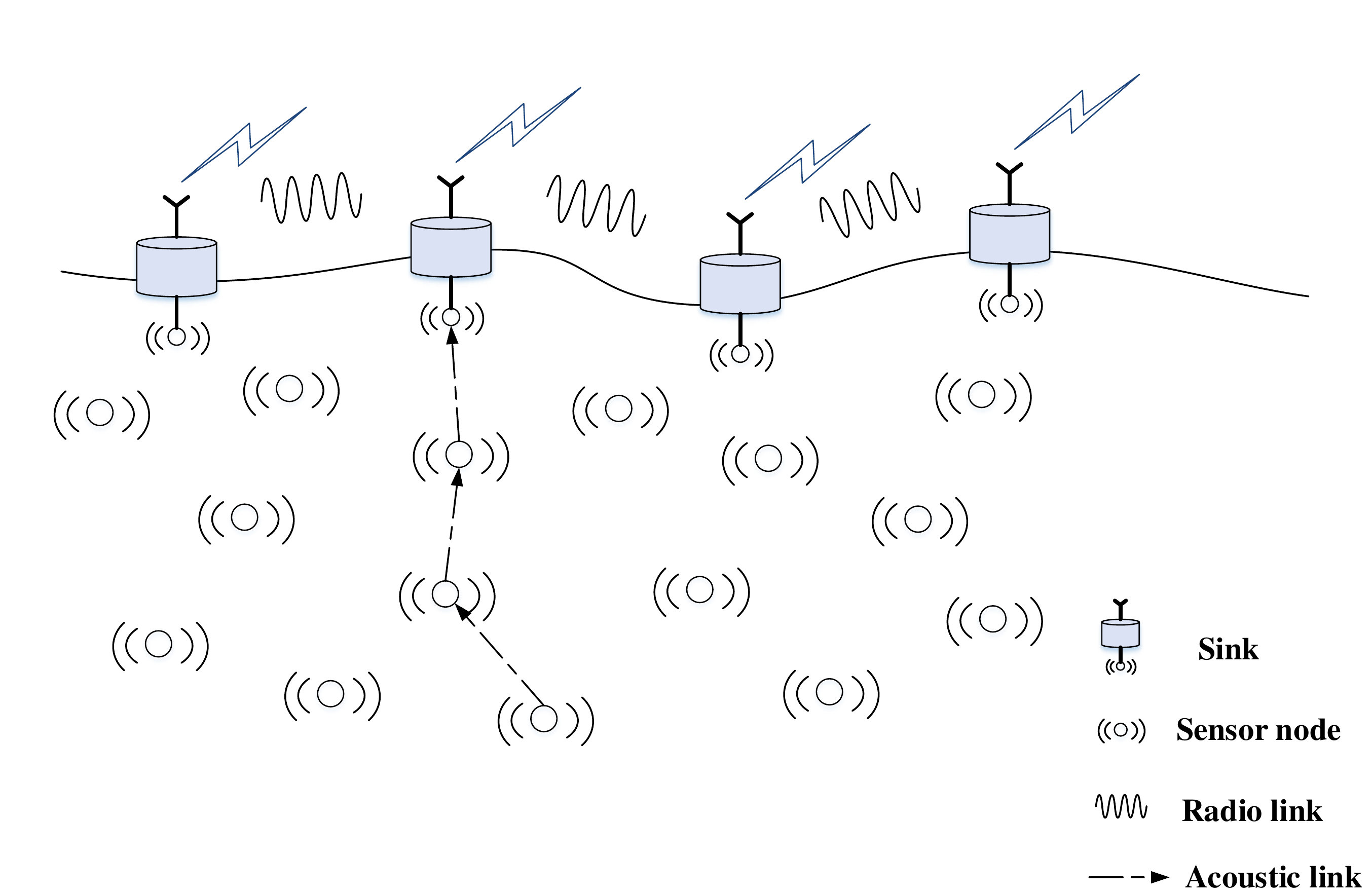}}
\caption{Multiple-sink underwater sensor network architecture.}
\label{fig2}
\end{figure}

\section{System Model}
In this section, we first introduce the network scenario, and then model the routing problem in UWSNs in the general framework of Q-leaning.

\subsection{Network Scenario}
In this paper, we consider a multiple-sink underwater sensor network architecture. An example of such UWSN is shown in Fig.~\ref{fig2}. The network consists of a set $N=SN\cup SK$ of nodes with a maximal transmission range of $R$, where $SN$ denotes the set of underwater sensor nodes, and $SK$ represents the set composed of sink nodes.

The underwater sensor nodes $SN=\left\{ s_1, s_2, s_3, \dots, s_{\left| SN \right|}\right\}$ are randomly deployed in 3-D area of interest, equipped with the acoustic modems and sensing devices to carry out 4-D (space and time) monitoring tasks. While the sinks $SK=\left\{ s_{\left| SN \right|+1}, s_{\left| SN \right|+2}, s_{\left| SN \right|+3}, \dots, s_{\left| SN \right|+\left| SK \right|}\right\}$ are deployed on the water surface, equipped with radio-frequency (RF) and acoustic modems. Acoustic channels are used for the underwater communication (i.e., the communication between underwater sensor nodes as well as between underwater sensor nodes and sinks), and radio channels are used for the maritime communication. Underwater sensor nodes collect data from monitoring areas and delivery the data to sinks which are considered as the destinations of underwater data packets. Sinks are located on the water surface; they send the received data to satellites with radio channel, and the satellites then transmit the data to the onshore data centers. Since the sinks can communicate to each other via radio channels efficiently, we assume that a packet is delivered successfully if it arrives at any of sink on the water surface.

\subsection{Q-learning Framework for UWSNs}
Q-learning is a value-based reinforcement learning (RL) method \cite{57}. Unlike traditional machine learning algorithms, reinforcement learning involves learning what to do©¤--how to map states to actions, so as to maximize a numerical reward signal. Using the RL, a system can achieve a goal in decision-making processes via its experience learning from interaction.

A RL task satisfied the Markov property can be considered as a MDP (Markov decision process) \cite{58}.  A particular MDP corresponds to a tuple $<\mathcal{S}, \mathcal{A}, \mathcal{P}, \mathcal{R}>$, where $\mathcal{S}$ is the states set, $\mathcal{A}$ is the actions set, $\mathcal{P}$ is the set of  transition probabilities, and $\mathcal{R}$ is the rewards set \cite{59}. In RL, a direct reward $r_{t+1}$ is defined to evaluate the return after taking an action at time $t$. And a policy $\pi$ is a mapping operation defined as $\pi:(s,a)\mapsto\pi(a\big|s)$, where $\pi (a\big|s)$ is the probability of taking action $a$ in state $s$.

Then we can define two value functions, the V-value and the Q-value. Informally, the V-value for state $s$, denoted as $V_{\pi}(s)$, is the expected cumulative reward when starting in state $s$ and following policy $\pi$ thereafter. We can define $V_{\pi}(s)$ as
\begin{equation}\label{1}
V_{\pi}(s)\triangleq \mathbb{E}_{\pi}[\sum_{i=0}^{+\infty}\gamma^ir_{t+i+1}\big|S_t=s],
\end{equation}
where $\mathbb{E}_{\pi}[\cdot]$ denotes the expected value following policy $\pi$. $\gamma\in[0,1]$ is defined as a discount factor.

Similarly, we can define the Q-value $Q_{\pi}(s,a)$ for a state-action pair $(s,a)$, which represents the expected cumulative reward of taking the action $a$ in state $s$ following policy $\pi$:
\begin{equation}\label{2}
Q_{\pi}(s,a)\triangleq \mathbb{E}_{\pi}[\sum_{i=0}^{+\infty}\gamma^ir_{t+i+1}\big|S_t=s,A_t=a].
\end{equation}

In Q-learning, the Q-value can be approximated recursively as follows:
\begin{equation}\label{3}
Q_{k+1}(S_t,A_t)\leftarrow\alpha[(r_{t+1})_{k+1}+\gamma V_{k}(S_{t+1})]+(1-\alpha)Q_{k}(S_t,A_t),
\end{equation}
where $\alpha\in(0,1]$ represents the learning rate, which reflects the rate of updating Q-value; $k$ denotes the iteration number. $V_{k}(S_{t+1})=\mathop{\max}_{a\in \mathcal{A}}Q_{k}(S_{t+1},a)$ is the approximated V-value of a given state $S_{t+1}$. For the sake of understanding (3), let us consider an example. Given that following policy $\pi$, we have performed a reinforcement learning task $k$ times and gotten the corresponding experiences, namely the V-value of each state and the Q-value of each state-action pair. Then in the $(k+1)$-th iteration, the Q-value in each timestep can approximated by (4) as follows.
\begin{equation}\label{4}
\begin{split}
&Q_{k+1}(S_t,A_t)\\
&=\alpha[(r_{t+1})_{k+1}+\gamma V_k(S_{t+1})]+(1-\alpha)Q_k(S_t,A_t)\\
&=\alpha[(r_{t+1})_{k+1}+\gamma \mathop{\max}_{a\in \mathcal{A}}Q_k(S_{t+1},a)]+(1-\alpha)Q_k(S_t,A_t).
\end{split}
\end{equation}

In UWSNs routing problems, we regard the entire underwater sensor network as a learning system. When a node $s_i$ is about to send a data packet, the system state is defined as $s_i$. Let $a_{s_j}$ be the action that the packet is sent to node $s_j$. All the states and actions in the UWSN make up the states set $\mathcal{S}$ and actions set $\mathcal{A}$. We consider the routing path as the policy $\pi$ in Q-learning, because a selected routing path can direct packet forwarding $a$ which is regarded as an action.

\section{QLFR Algorithm}
In this section, we first give an overview of the proposed QLFR algorithm and then describe it in detail, including the reward functions, a new holding time mechanism and a multipath suppression scheme.

\subsection{QLFR Overview}
In QLFR, before sending a packet, the sender should make a routing decision for forwarding the packet. In order to find the optimal forwarder, QLFR uses the Q-learning technique to compute the Q-values of all its neighboring nodes. A node with a larger Q-value is considered to be better suited for forwarding packets than one with a lower Q-value. Then the sender selects the neighbors with a smaller depth than itself among all its neighboring nodes, and sorts the selected nodes in accordance with their Q-values. Next, the sender creates a priority list where nodes are sorted in descending order of their Q-values, and embeds the priority list in the sending packet. When a node receives a packet, it will retrieve the priority list. If its ID is in the priority list, the node will hold the packet for a period of time dubbed the holding time. If not, it will simply drop the packet. A node with a higher priority will have a shorter holding time, and thus it will send packets earlier than its peers. For the other nodes, during their holding time, if they overhear the packet transmitted by a node with a higher priority, they will give up  forwarding this packet.

\begin{figure*}[b]
\hrulefill
\begin{align*}
Q_{k+1}(s_i,a_j)&=\alpha[(r_{s_is_j}^{a_j})_{k+1}+\gamma{\max}_{a\in \mathcal{A}}Q_k(s_j,a)]
+(1-\alpha)Q_k(s_i,a_j)\\
&=\alpha[(r_{s_is_j}^{a_j})_{k+1}+\gamma V_k(s_j)]+(1-\alpha)Q_k(s_i,a_j).
\tag{8}
\end{align*}
\end{figure*}

In QLFR, a data packet is delivered from a source to a sink based on the depth information. During this process, the depth of forwarding nodes decreases, while the data packet approaches the sink node. To balance energy consumption, QLFR takes into account the residual energy of nodes during the packet forwarding process. That is to say, QLFR tries to choose the neighbors with smaller depths and more residual energy to forward a packet. On the other hand, for each sender, there may exist more than one eligible node to forward a packet. If all the eligible nodes attempt to forward the same packet, it will cause both high energy consumption and a high collision rate. Therefore, in order to minimize energy consumption as well as the collision rate, one should carefully choose the forwarding nodes. Towards this end, QLFR designs a new Q-learning based holding time mechanism to schedule the forwarding of packets.
\subsection{Reward Function}
In QLFR, we assign the priority for forwarding candidate nodes according to the Q-value during the data packet transmissions. As we discussed in Section III, Q-value is defined as the expected cumulative reward, so how to define the direct reward is crucial to our proposed QLFR algorithm.

To extend the network lifetime and reduce latency, we take both the residual energy and depth information of sensor nodes into consideration. Consider the scenario that a data packet is forwarded from node $s_i$ to $s_j$, the reward function is defined as:
\begin{equation}
\setcounter{equation}{5}
r_{s_is_j}^{a_j}=-c_e(s_i)-c_e(s_j)-c_d(s_i,s_j),
\end{equation}
where $c_e(\cdot)$ is a residual energy-related cost function and $c_d(\cdot)$ is a depth-related cost function. In our algorithm, after a sender transmit a data packet, it will receive a negative reward. Both of these two cost functions are in the range of [0,1].

Considering the residual energy of node $s_i$ in packets forwarding, the residual energy-related reward function $c_e(s_i)$ is define as:
\begin{equation}\label{6}
c_e(s_i)=1-\frac{e_{\rm res}(s_i)}{e_{\rm ini}(s_i)},
\end{equation}
where $e_{\rm res}(s_i)$ is the residual energy of node $s_i$ and $e_{\rm ini}(s_i)$ is its initial energy. Since all sensor nodes have the same initial energy $e_{\rm ini}(s_i)$, a sensor node with the less residual energy has a higher cost $c_e(s_i)$. A node with higher cost is more reluctant to involve in the communications. By considering the residual energy, sensor nodes are used in a relatively fair way, which implies energy balanced.

$c_d(s_i,s_j)$ is the depth-related cost function, which lies on the difference between the depth of sender $s_i$ and its neighbor $s_j$. It is based on the fact that a node with a smaller depth is closer to sink nodes. So $c_d(s_i,s_j)$ can describe the end-to-end delay, and we define it as:
\begin{equation}\label{7}
c_d(s_i,s_j)=\frac{1}{2}\Bigg(1-\frac{d(s_i,s_j)}{\left| d \right|_{\rm max}}\Bigg),
\end{equation}
where $d(s_i,s_j)$ is the difference between the depth of node $s_i$ and its one hop neighboring node $s_j$ (i.e., $d(s_i,s_j)=depth(s_i)-depth(s_j)$), $\left| d \right|_{\rm max}$ denotes the maximum of the absolute value of $d(s_i,s_j)$. Clearly, a forwarding candidate node with the larger depth is further away from the destination, thus its corresponding $d(s_i,s_j)$ will be smaller, which cause a higher cost $c_d(s_i,s_j)$.

By the definition of reward functions above, we can get the corresponding Q-value in different packet transmission rounds via an iterative way according to (3). A packet transmission round means that the packet is routed from the source node to any of the sink nodes. Specifically, the Q-value in $(k+1)$-th round packet transmission can be calculated as (8), presented at the bottom of this page.

In QLFR, before transmitting a data packet, the sender creates a priority list for its next hop forwarding candidates according to the Q-value, and embeds the priority list into the data packet. The sequence number of a candidate node in the priority list, denoted as $n$, represents its priority level. For example, the first node in the priority list, namely $n=1$, means that it has the highest priority; $n=2$ means the node has the next highest priority, and so on. After receiving the data packet, a node first retrieve the priority list. If the node finds its own ID in the priority list, it will perform according to the proposed holding time mechanism based on its priority level, which will discuss elaborately in next part; otherwise, it will discard the packet.
\begin{figure}[t]
\centerline{\includegraphics[width = 1\columnwidth]{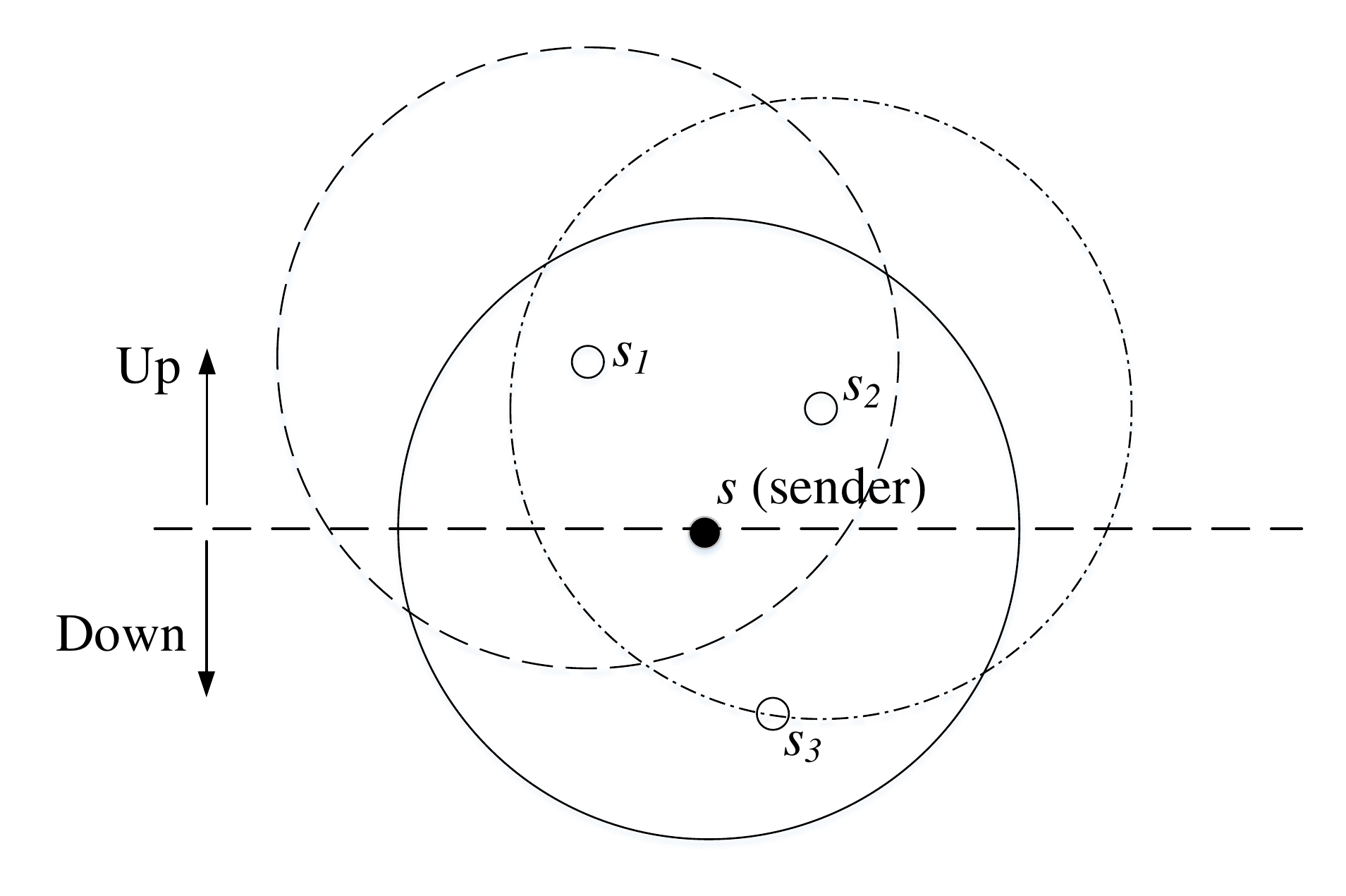}}
\caption{Example of the holding time mechanism.}
\label{fig3}
\end{figure}

\subsection{Holding Time}
As mentioned earlier, in our routing protocol, holding time is used to schedule packet forwarding. For node $s$, the holding time is calculated based on the sequence number $n$ in the priority list, which can represent the priority of node $s$. Nodes with different priorities will have different holding time. To select a better routing and reduce redundant transmissions, QLFR tries to choose the node with the higher priority to forward a data packet, meanwhile, prevent other nodes with lower priorities from forwarding the same packet.

Based on the above analysis, the holding time can be expressed using a linear function of $n$ as follow.
\begin{equation}
\setcounter{equation}{9}
\tau(n)=kn+b.
\end{equation}

An example of the holding time mechanism is given in Fig.~\ref{fig3}. When $s$ sends a data packet, $s_1$, $s_2$ and $s_3$ will receive this data packet, since they are all neighbors of $s$. Node $s_3$ is below $s$, so it drops the data packet. Node $s_1$ and $s_2$ are both qualified candidates. Assuming that $s_1$ receives the packet at time $t_1$, $s_2$ receives the packet at time $t_2$, the propagation delay between $s_1$ and $s_2$ is denoted as $t_{\rm prop}$.
We assume that $Q(s, s_1)> Q(s, s_2)$. Thus, $s_1$ is preferred to forward the packet; $s_2$ will give up forwarding if it overhears the packet sent by $s_1$ before it forwards the packet on schedule. Let $n_1$ and $n_2$ be the sequence numbers of nodes $s_1$ and $s_2$ in the priority list, respectively.

The difference of holding times of two adjacent nodes need to be long enough so that lower priority node can hear the forwarding of the higher priority node before the lower priority node forwarding the packet on schedule. Therefore, the following two constraints should be satisfied:
\begin{equation}\label{10}
\tau(n_1)\textless\tau(n_2),
\end{equation}
\begin{equation}\label{11}
t_1+\tau(n_1)+t_{\rm prop}\leq t_2+\tau(n_2).
\end{equation}

Substituting $\tau(n)=kn+b$ into the above constraints (10) and (11), we have
\begin{equation}\label{12}
k\geq \frac{t_1+t_{\rm prop}-t_2}{n_2-n_1},(k\textgreater0).
\end{equation}
Here, $k$ is positive. As long as the above inequation (12) holds, constraints (10) and (11) can be both satisfied. Let $v_0$ denote the speed of acoustic waves in water and $R$ denote the maximal transmission range of a sensor node, the maximal propagation delay of one hop is represented by $t_{\rm max}=\frac{R}{v_0}$. Thus, $t_1-t_2\leq\frac{R}{v_0}$ and $t_{\rm prop}\leq\frac{R}{v_0}$. Substituting them to (12), we have
\begin{equation}\label{13}
\begin{split}
\frac{t_1+t_{\rm prop}-t_2}{n_2-n_1}\leq\frac{2\cdot\frac{R}{v_0}}{n_2-n_1}=\frac{2t_{\rm max}}{n_2-n_1}.
\end{split}
\end{equation}

Let's set $k=\frac{2t_{\rm max}}{h}$, $h\in\mathbb{N^*}$. When $n_2-n_1\geq h$, inequation (12) will hold and we can guarantee that node $s_1$ can send a packet earlier than $s_2$ and prevent $s_2$ to forward the same packet.

In QLFR, the first node in the priority list has the highest priority. For reducing end-to-end delay, the holding time of first node in the priority list is set to zero. Thereby, we can have:
\begin{equation}\label{14}
\begin{split}
\tau(1)&=k\cdot1+b=0\\
b&=-k.
\end{split}
\end{equation}

Substituting $k$ and $b$ into linear function $\tau(n)=kn+b$, we can compute the expression of holding time as follow:
\begin{equation}\label{15}
\tau(n)=k\cdot(n-1)=\frac{2t_{\rm max}}{h}(n-1),(h\in\mathbb{N^*}).
\end{equation}

If we choose a large $k$, nodes will have longer holding times. This may cause longer end-to-end delays. Simultaneously, those nodes with lower priority is more likely to be suppressed by the nodes with higher priority. Therefore, it can reduce the redundant transmission, and result in lower energy consumption. On the contrary, if $k$ is set to be a small value, nodes have shorter holding times, it may lead to shorter end-to-end delays but higher energy consumption.
\begin{figure}[t]
\centerline{\includegraphics[width = 1\columnwidth]{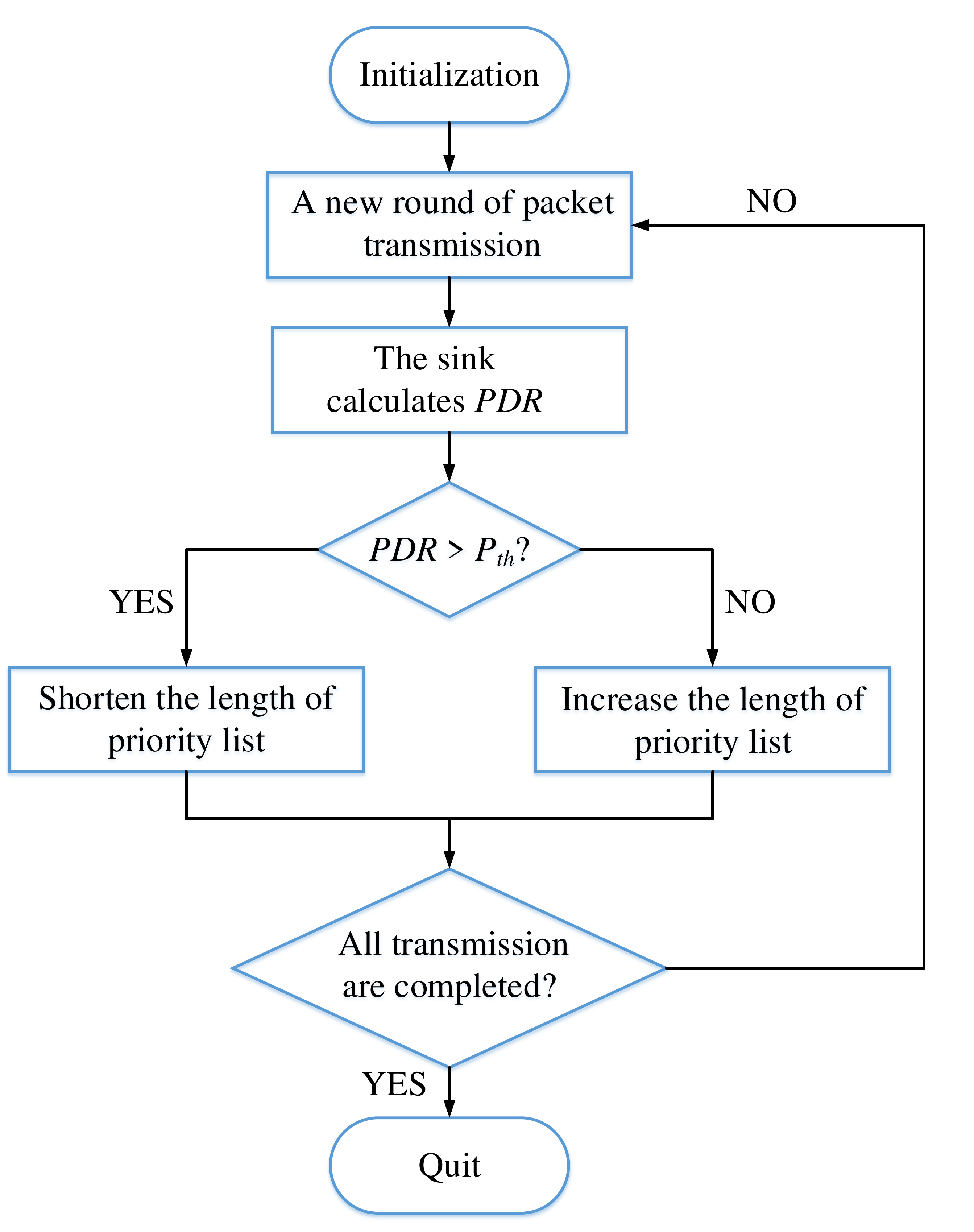}}
\caption{Flowchart of the proposed multipath suppression scheme.}
\label{fig4}
\end{figure}

\subsection{Multipath Suppression Scheme}
In practice, we tend to increase the value of $h$ appropriately to ensure a small end-to-end delay, but it inevitably leads to redundancy. To achieve higher energy efficiency, we should further restrict the packet transmissions.

However, suppressing the packet transmissions excessively will reduce the packets delivery ratio (PDR). In some specific scenarios, such as military communications, packet delivery ratio should take priority over energy efficiency. Therefore, to improve the energy efficiency while ensure a high packet delivery ratio, we propose a new multipath suppression scheme. The flowchart of the proposed scheme is shown in Fig.~\ref{fig4}.

The length of the priority list is initialized to a constant integer. During packets forwarding, the source attaches the total number of generated packets to the sending packet. When the sink node receives the packet, it can calculate the packet delivery ratio via dividing the successfully received data packets to total generated packets.

If the delivery ratio is higher than the threshold set according to the application scenario, the length of the list will be shortened during the next packet forwarding round to improve the energy efficiency.

If the delivery ratio is lower than the threshold, sink node will inform the source node through broadcasting a message of increasing the length of the priority list. Then the source node will embed the message to next data packet, and the nodes who eligible to forward the packet will increase the length of the list according to the message to increase the delivery ratio.

\section{Routing Protocol Design}
In this section, we elaborate our routing protocol from three aspects, involving packet structure, routing knowledge exchange and data packet forwarding.

\begin{figure}[t]
\centerline{\includegraphics[width = 1\columnwidth]{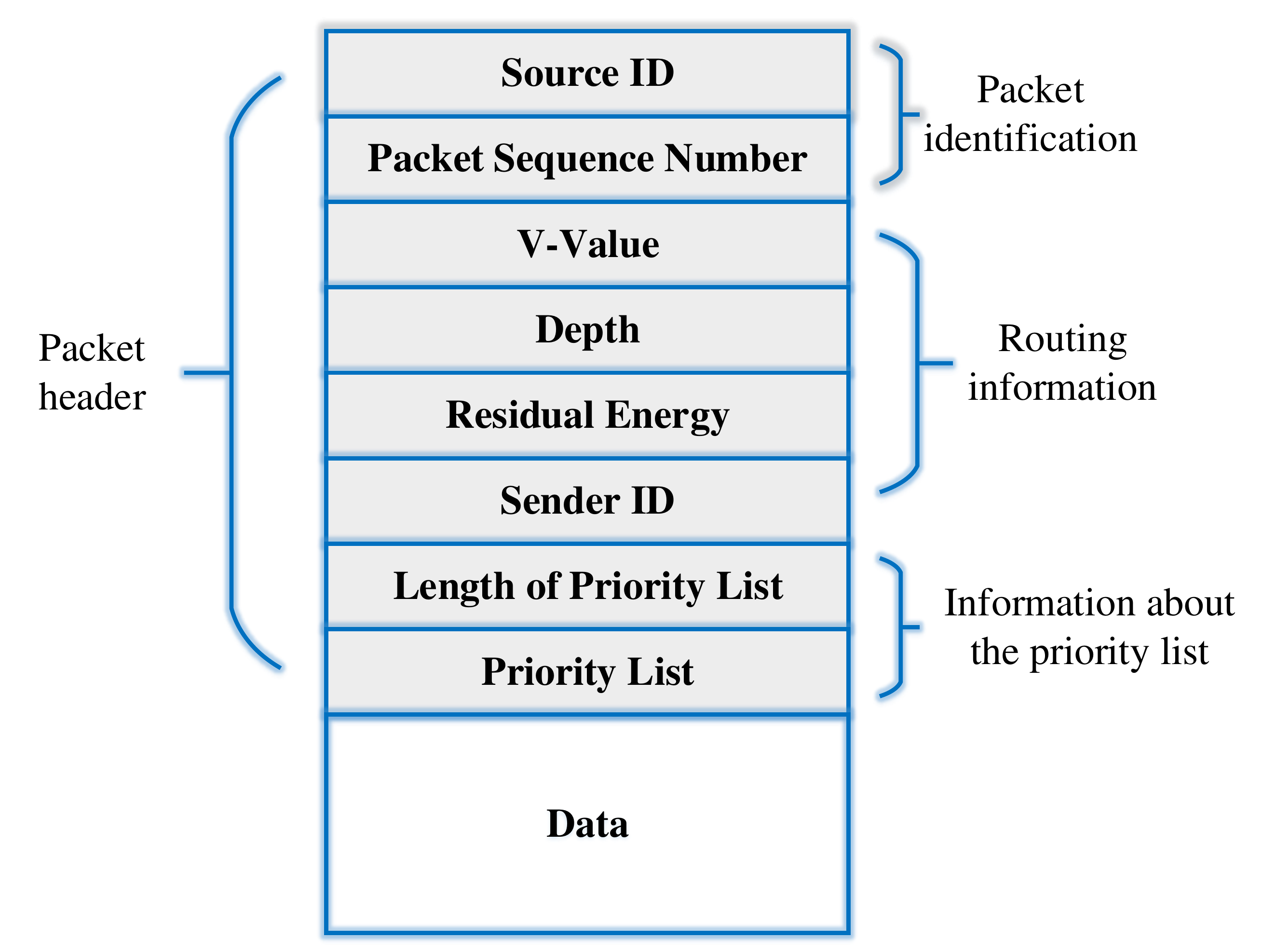}}
\caption{Packet structure.}
\label{fig5}
\end{figure}

\subsection{Packet Structures}
The packet structure in the network is illustrated in Fig.~\ref{fig5}. The packet header includes three parts: packet identification fields, routing information fields and the information about the priority list.

The packet identification fields are
\begin{itemize}
\item \textbf{\emph{Source ID}}, the identifier of the source node,
\item \textbf{\emph{Packet Sequence Number}}, the unique ID of the packet.
\end{itemize}
They are depended on the source nodes of the packet. These two fields are used to differentiate packets in data forwarding. Generally, they are permanent during the entire lifetime of the packet.

Before sending the packet, a node should embed its information in these fields, which include
\begin{itemize}
\item \textbf{\emph{V-Value}}, the V-value of the current node,
\item \textbf{\emph{Depth}}, the depth information of the current node,
\item \textbf{\emph{Residual Energy}}, the residual energy information of the current node,
\item \textbf{\emph{Sender ID}}, the ID or address of the current node.
\end{itemize}
Upon receiving a data packet, every node retrieves these fields from packet header and updates its neighbors' information with the latest routing information which helps them making optimal routing decisions.

The information about the priority list is used to advance packet and to help the forwarding candidates cooperating with each other, which are
\begin{itemize}
\item $\textbf{\emph{Length of Priority List}}$, the number of nodes eligible to forward data packets,
\item $\textbf{\emph{Priority List}}$, routing decision made by the sender.
\end{itemize}
Length of Priority is used to control the number of forwarding candidates as mentioned above. After making a routing decision, the sender puts the priority list of the forwarding candidate nodes to the field of Priority List. According to the priority list, the next-hop nodes who received the packet will decide to perform the forwarding or to drop the packet.

Other than the packet header, the \textbf{\emph{Data}} is optional. This part is the message that should be sent to the destination. If \textbf{\emph{Data}} is absent, the packet is used only for routing information exchanging, which will be described in the next part.

\subsection{Routing Knowledge Exchange}
In order to make an optimal routing decision, sensor nodes should know their neighbors' V-values $V(s_i)$, depth $depth(s_i)$ and residual energy $e_{\rm res}(s_i)$ for calculating the Q-values. A tuple expressed as $<V(s_i), depth(s_i), e_{\rm res}(s_i)>$ is used to represent the routing knowledge of a sensor node. The routing protocol applies a distributed mechanism, that is, all the sensor nodes send the routing knowledge to their neighbors. There are two methods for routing knowledge exchange:

\begin{enumerate}
\item[1)]The routing knowledge can be exchanged \textbf{\emph{simultaneously with data packets transmission}}. When transmitting a data packet, routing knowledge will be attached to it. A node is able to acquire its neighbors' routing knowledge from the incoming data packets.

    \item[2)]The routing knowledge can be exchanged by \textbf{\emph{a Hello packet containing only routing knowledge}}. Each node in UWSNs will broadcast a Hello packet periodically. This packet has no payload, and it is only used to exchange the routing knowledge. This kind of broadcasts is considered as a complementary method for routing information exchange. Since every node can acquire the routing information via data packets, we do not need broadcast special control packets. The period of broadcasting Hello packets can be set very long, and thus, it is reasonable to ignore the overhead of this part.
\end{enumerate}

\subsection{Data Packet Forwarding}
Based on the crucial components of the proposed routing protocol defined above, the procedure of data packet forwarding in QLFR is discussed in this part. This procedure is summarized in Algorithm ~\ref{Algorithm 1}.

When a sender prepares to transmit a data packet, it first calculates the Q-values associated with each of its neighboring nodes using the acquired routing knowledge. A priority list is formed by opting the neighboring nodes simultaneously having two characteristics: 1) the nodes are with smaller depth than the sender and 2) with large Q-values. $\textbf{\emph{Length of Priority List}}$ controls the number of forwarding candidates priority list. The neighboring nodes in the priority list are regarded as the candidates for the next hop. Before sending the packet, the sender updates the packet header with its own information and the priority list of the next hop candidates.

\begin{algorithm}[t]
  \caption{Algorithm for data packet forwarding in QLFR. $K$ is the data packet. $s_i$ is the node that currently receives the data packet. $\tau$ denotes the holding time of $s_i$ to hold the packet. $Neighbor(s_i)$ is the set included all the neighboring nodes of $s_i$. $s_j$ is an element in the set $Neighbor(s_i)$. }
  \label{Algorithm 1}
  \begin{algorithmic}[1]
  \STATE Onhearing $K$
  \STATE Get the sender's information from the header of $K$
  \STATE Get the priority list from the header of $K$
  \IF {$s_i\notin$ priority list}
  \STATE Drop $K$
  \ELSIF {$s_i$ has forwarded $K$}
  \STATE Drop $K$
  \ELSE
  \STATE Calculate $\tau$
  \STATE Calculate $Q(s_i,s_j)$, $s_j \in Neighbor(s_i)$
  \STATE Set the new priority list for next hop candidates
  \IF {$s_i$ overhears $K$ during $\tau$}
  \STATE Drop $K$
  \ELSE
  \STATE Update the header of $K$
  \STATE Send $K$ when $\tau$ expires
  \ENDIF
  \ENDIF
  \end{algorithmic}
\end{algorithm}

Upon receiving a packet, no matter whether or not a node is designated as the qualified forwarder, it extracts the routing knowledge of the sender from the packet header, and updates the corresponding neighbor information. If the node is not eligible to forward the packet, it simply drops the packet; otherwise, the node will check whether it has forwarded the data packet before.

In this case, if the packet has been forwarded by the node before, it is discarded by the node, hence, other candidate nodes will have the opportunity to forward this packet. Otherwise, the node calculates the holding time for the packet based on its sequence number in the priority list. The first node in the priority list means that it has the highest priority, and it forwards the packet immediately without waiting; while other nodes in the priority list hold the packet for the calculated holding time. During the holding time, if a node overhearing the same packet, it will give up the forwarding of this packet, as another node with higher priority has already forwarded the packet; if not, the node will transmit the packet when holding time expires.

\section{Theoretical Analysis}
In this section, we first calculate the delivery probability between two nodes, which is defined as the successful probability of a packet transmitted from one node to another in our protocol, and is crucial in deriving the other performance metrics. Then, we analyze the performance of our proposed routing protocol with respect to four performance metrics, i.e., the packet delivery ratio, end-to-end delay, energy consumption and network lifetime.

\subsection{Delivery Probability}
The Urick's model \cite{62}, which is a commonly used model in the UWSN, is adopted to formulate the underwater acoustic channel in this paper. In this model, the attenuation of the underwater acoustic signal with frequency $f$ (in kHz) at transmission distance $l$ (in meter) is given by
\begin{equation}\label{16}
A(l,f)=A_0l^{\kappa}a(f)^l.
\end{equation}
where $A_0$ is a constant attenuation factor, which models the signal attenuation caused by the propagation effects, such as scattering, refraction and multipath propagation. $\kappa \in [1,2]$ is the spreading loss factor. $a(f)$ denotes the absorption coefficient and can be calculated by the Thorpe formula \cite{63} as follows
\begin{equation}\label{17}
10\log a(f)=2.75\times10^{-4}f^2+\frac{44f^2}{4100+f}+\frac{0.11f^2}{1+f^2}+10^{-3}.
\end{equation}

Then, for an underwater acoustic link with signal frequency $f$ and transmission distance $l$, the average signal-to-noise ratio (SNR) at the receiver of this link can be shown as
\begin{equation}\label{18}
\overline{SNR(l,f)}=\frac{e_b/A(l,f)}{N_0}=\frac{e_b}{N_0 A_0l^{\kappa}a(f)^l},
\end{equation}
where $N_0$ represents the power density of the noise modeled as the additive white Gaussian noise (AWGN),  and $e_b$ is the transmit energy per bit, which are both constants. In addition, Rayleigh fading is adopted to simulate small scale fading \cite{64,65,66}, in which the probability density of the SNR can be described as follows
\begin{equation}\label{19}
f_{SNR}(l,f,X)=\frac{1}{\overline{SNR(l,f)}}e^{-\frac{X}{\overline{SNR(l,f)}}}.
\end{equation}

Thereby, we can derive the probability of data transmission errors per bit using the following formula
\begin{equation}\label{20}
p_e(l,f)=\int_{0}^{\infty} p_e(X)f_{SNR}(l,f,X)\, dX,
\end{equation}
where $p_e(X)$ represents the probability of data transmission errors using any selected modulation scheme at a SNR of $X$. Furthermore, similar to \cite{67,68,69}, the binary phase shift keying modulation is employed in this paper. Thus, as in \cite{70,71}, the corresponding probability of data transmission errors per bit can be shown as
\begin{equation}\label{21}
p_e(l,f)=\frac{1}{2}\Bigg(1-\sqrt{\frac{\overline{SNR(l,f)}}{1+\overline{SNR(l,f)}}}\Bigg).
\end{equation}

Therefore, letting the signal frequency be $f$ kHz, the transmission distance be $l$ meters and the data packet size be $M$ bits, the delivery probability can be calculated as follows
\begin{equation}\label{22}
p(l,f,M)=(1-p_e(l,f))^M.
\end{equation}

\subsection{Expected Packet Delivery Ratio}
The packet delivery ratio describes the probability that data packets are successfully forwarded from the source node to the sink. In order to analyze the packet delivery ratio, we first discuss the delivery probability of one hop. Let us assume that the current sender is $s_i$, and $Neighbor(s_i)$ is the set denoting all its neighboring nodes. In our proposed routing protocol, $s_i$ selects a subset of $Neighbor(s_i)$ to form the forwarding candidate set $\varphi(s_i)$ and creates a priority list for these selected candidates according to their Q-values. In the priority list $(s_1, s_2, \dots, s_{j-1}, s_j, \dots)$, candidates are sorted in descending order of their priority levels. That is, the front candidates have higher priorities. According to (22), when the signal frequency is $f$ kHz, the transmission distance is $l$ meters and the data packet size is $M$ bits, the delivery probability between sender $s_i$ and a candidate node $s_j$ can be given as follows
\begin{equation}\label{23}
p_{s_is_j}(l,f,M)=(1-p_e(l,f))^M.
\end{equation}

For brevity of exposition, $p_{s_is_j}(l,f,M)$ is simply denoted by $p_{s_is_j}$ in the remaining of this section. Moreover, a coordination scenario \cite{72} among these selected candidates is considered in our analysis. In this scenario, if a candidate $s_j$ is about to forward a data packet, then the following two conditions should be satisfied:
\begin{itemize}
\item the packet is successfully sent from sender $s_i$ to the forwarding candidate node $s_j$; and
\item transmission errors have occurred when the packet is forwarded by candidate nodes with higher priorities than node $s_j$.
\end{itemize}

Therefore, the probability of the packet successfully forwarded by candidate node $s_j$ can be calculated as
\begin{equation}\label{24}
P_{s_is_j}=p_{s_is_j}\prod_{k=1}^{j-1}(1-p_{s_is_k}),
\end{equation}
where $p_{s_is_j}$ is the delivery probability between sender $s_i$ and candidate node $s_j$ defined in (23), and $\prod_{k=1}^{j-1}(1-p_{s_is_k})$ describes the probability that transmission errors have occurred between the sender $s_i$ and the forwarding candidate nodes with higher priorities than candidate $s_j$.

A packet is successfully delivered in one hop means that the packet is correctly delivered from sender $s_i$ to any one of its next-hop forwarding candidates (i.e., $\forall s_j\in\varphi(s_i)$). Therefore, the delivery probability of one hop in our proposed routing protocol can be calculated as follows
\begin{equation}\label{25}
P_{\rm one-hop}=\sum_{\forall s_j\in\varphi(s_i)}P_{s_is_j}.
\end{equation}

The packet is successfully routed from $s_i$ to the sink node means that the packet is correctly forwarded in each hop. Thereby, we can calculate the corresponding delivery probability from $s_i$ to the sink node in a recursive manner. More specifically, we first calculate the probability of the packet forwarded by each of $s_i$'s candidate nodes. Then, the delivery probability from a forwarding candidate node to the sink node should be calculated in the same recursive way. This process can be formulated as
\begin{equation}\label{26}
P_{s_i-\rm sink}=\sum_{\forall s_j\in\varphi(s_i)}P_{s_is_j}P_{s_j-\rm sink},
\end{equation}
where $P_{s_is_j}$ denotes the probability that the packet is transmitted from $s_i$ and forwarded by $s_i$'s candidate node $s_j$ in the next hop as defined in (24). $P_{s_i-\rm sink}$ and $P_{s_j-\rm sink}$  represent the delivery probabilities from $s_i$ and $s_j$ to the sink node, respectively. In addition, when $s_j$ is a sink node, $P_{s_j-\rm sink}$ is set to be 1. This process is iterated from the sink node to the source node, and then the expected packet delivery ratio can be calculated.

\subsection{Expected End-to-End Delay}
End-to-end delay describes the duration of a packet being routed from the source node to the sink node, which is also a crucial quantitative metric for evaluating the performance of routing protocols. Similar to the analysis of the packet delivery ratio, we first discuss the delay of one hop. In our proposed routing protocol, the delay in one hop consists of two parts:
\begin{itemize}
\item the holding time of the sender, which is the duration that the sender should wait before transmitting the packet; and
\item the latency caused by the propagation of the packet from the sender to its next-hop forwarding candidates.
\end{itemize}

We consider node $s_i$ as the current sender. The holding time of $s_i$ in our proposed routing protocol is calculated based on its priority level. Thus, before $s_i$ transmits the packet, the expected holding time can be calculated as
\begin{equation}\label{27}
\begin{split}
\tau_{s_i}^{\rm expected}&=\sum_{\left\{ s_k| s_i\in \varphi(s_k) \right\}}\tau_{s_is_k} P_{s_ks_i} \\
&=\sum_{\left\{ s_k| s_i\in \varphi(s_k) \right\}}\tau_{s_is_k}  p_{s_ks_i}\prod_{m=1}^{i-1}(1-p_{s_ks_m}),
\end{split}
\end{equation}
where $s_k$ is one of $s_i$'s the neighboring nodes and acts as the sender of $s_i$. $\tau_{s_is_k}$ represents the holding time that $s_i$ should wait before transmitting the data packet when the packet is sent by $s_k$, which is defined in (15). $P_{s_ks_i}$ denotes the probability that the packet is transmitted from $s_k$ and forwarded by $s_i$ as defined in (24).

To calculate the propagation time of the data packet from $s_i$ to one of its next-hop forwarding candidates $s_j$, we first introduce the speed model of acoustic waves in water, which is a function of the depth (or hydraulic pressure), salinity and temperature of the water \cite{73,74}. This speed function can be modelled as follows
\begin{equation}\label{28}
\begin{split}
v_0=&-7.139\times 10^{-13}H^3T+2.374\times 10^{-2}T^3\\
&+1.675\times 10^{-7}H^2-5.304\times 10^{-2}T^2\\
&-1.025\times 10^{-2}T(S-35)+0.163H\\
&+4.591T+1.34(S-35)+1448.96,
\end{split}
\end{equation}
where $H$ (in meter) is the underwater depth, $T$ (in degree celsius) denotes the temperature, and $S$ (in part per thousand) represents the salinity of the water. In general, for the sake of simplicity, this propagation speed is approximated to be a constant and set to $v_0=1500$ m/s \cite{75,76,77}, which is also adopted in this paper. Thus, the propagation time of a packet from $s_i$ to one of its next hop forwarding candidates $s_j$ is
\begin{equation}\label{29}
t_{s_is_j}=\frac{D_{s_is_j}}{v_0},
\end{equation}
where $D_{s_is_j}$ is the distance between the sender $s_i$ and its forwarding candidate $s_j$. Thereby, the latency caused by the transmission from $s_i$ to $s_j$ can be given by
\begin{equation}\label{30}
\begin{split}
T_{s_is_j}&=\tau_{s_i}^{\rm expected}+t_{s_is_j}\\
&=\sum_{\left\{ s_k| s_i\in \varphi(s_k) \right\}}\tau_{s_is_k}P_{s_ks_i}+\frac{D_{s_is_j}}{v_0}.
\end{split}
\end{equation}

Similar to the derivations of the packet delivery ratio, the expected end-to-end delay from $s_i$ to the sink node is also derived recursively. We first calculate the delay between $s_i$ and each of its forwarding candidates, and then the expected end-to-end delay from the forwarding candidate to the sink node should be calculated in the same recursive manner. We express the process as follows
\begin{equation}\label{31}
T_{s_i-\rm sink}=\sum_{\forall s_j\in\varphi(s_i)}(T_{s_is_j}+T_{s_j-\rm sink})P_{s_is_j},
\end{equation}
where $P_{s_is_j}$ denotes the probability that the packet is transmitted from $s_i$ and forwarded by $s_i$'s next hop candidate node $s_j$ as defined in (24). $T_{s_i-\rm sink}$ and $T_{s_j-\rm sink}$ represent the expected end-to-end delay from $s_i$ and $s_j$ to the sink node, respectively. Especially, if $s_j$ is a sink node, $T_{s_j-\rm sink}$ is equal to zero. The recursive process is performed from the sink to the source, and then the expected end-to-end delay is obtainable.

\subsection{Expected Energy Consumption}
Now, we analyze the energy consumption of node $s_i$. For a node in an underwater sensor network, its energy is consumed for two reasons, i.e., transmitting and receiving data packets. To analyze the energy consumed by $s_i$ for packet transmission, we should first know the outgoing traffic of $s_i$ (i.e., the packets forwarded by $s_i$), which can be also calculated in a recursive way. First, we should obtain the outgoing traffic of $s_i$'s sender, and then calculate the probability that the traffic is forwarded by $s_i$ afterwards.
\begin{equation}\label{31}
\begin{split}
\lambda_{s_i}&=\sum_{\left\{ s_k| s_i\in \varphi(s_k) \right\}}P_{s_ks_i} \lambda_{s_k}\\
&=\sum_{\left\{ s_k| s_i\in \varphi(s_k) \right\}} p_{s_ks_i}\prod_{m=1}^{i-1}(1-p_{s_ks_m}) \lambda_{s_k},
\end{split}
\end{equation}
where $s_k\in Neighbor(s_i)$ is a sender of $s_i$. $P_{s_ks_i}$ represents the probability that the packet is transmitted from $s_k$ and then forwarded by $s_i$, as defined in (24). $\lambda_{s_i}$ and $\lambda_{s_k}$ are the outgoing traffic of node $s_i$ and its sender $s_k$, respectively. Especially, for a source node, the outgoing traffic is the packets generated by itself.

After having obtained the outgoing traffic $\lambda_{s_i}$, one can calculate the packet transmission time $\delta_{s_i}^t$ as follows
\begin{equation}\label{33}
\delta_{s_i}^t=\lambda_{s_i} \frac{M}{\mu},
\end{equation}
where $M$ (in bit) is the size of a data packet and $\mu$ (in bps) denotes the data transmission rate. Therefore, given the packet transmission power $\Psi_t$, the energy consumption of $s_i$ for packet transmission (i.e., $E_{s_i}^t$) can be given as
\begin{equation}\label{34}
E_{s_i}^t=\delta_{s_i}^t \Psi_t.
\end{equation}

Similarly, in order to analyze the energy consumption caused by packet reception, it is necessary to calculate the duration of packet reception. Thanks to the broadcast nature of the underwater acoustic channel, node $s_i$ is able to overhear all the data packets transmitted from its neighboring nodes, even if the packets are not for itself. Thus, the amount of time that node $s_i$ spent to receive data packets is calculated by
\begin{equation}\label{35}
\delta_{s_r}^r=\sum_{s_j\in Neighbor(s_i)}\lambda_{s_j} \frac{M}{\mu}.
\end{equation}

The energy consumed by node $s_i$ for receiving packets is calculated by the power and the amount of time it spends to receive
\begin{equation}\label{36}
E_{s_i}^r=\delta_{s_i}^r \Psi_r,
\end{equation}
where $\Psi_r$ is the power in receiving data packets.

Taking both packet transmission and reception into consideration, we can calculate the energy consumption of node $s_i$ as follows
\begin{equation}\label{37}
\begin{split}
E_{s_i}&=E_{s_i}^t+E_{s_i}^r\\
&=\delta_{s_i}^t \Psi_t+\delta_{s_i}^r \Psi_r\\
&=\lambda_{s_i} \frac{M}{\mu} \Psi_t+\sum_{s_j\in Neighbor(s_i)}\lambda_{s_j} \frac{M}{\mu} \Psi_r.
\end{split}
\end{equation}

\subsection{Expected Network Lifetime}
Setting the network running time until now to be $T_{\rm run}$, the average energy consumption per second for node $s_i$ can be shown as
\begin{equation}\label{38}
\begin{split}
e_{s_i}&=\frac{E_{s_i}}{T_{\rm run}}\\
&=\frac{\lambda_{s_i} \frac{M}{\mu} \Psi_t+\sum_{s_j\in Neighbor(s_i)}\lambda_{s_j} \frac{M}{\mu} \Psi_r}{T_{\rm run}}.
\end{split}
\end{equation}

Therefore, the lifetime of node $s_i$ can be estimated as
\begin{equation}\label{39}
\begin{split}
\Gamma_{s_i}&=\frac{e_{\rm ini}(s_i)}{e_{s_i}}=\frac{e_{\rm ini}(s_i)}{\frac{E_{s_i}}{T_{\rm run}}}\\
&=\frac{e_{\rm ini}(s_i) T_{\rm run}}{\lambda_{s_i} \frac{M}{\mu} \Psi_t+\sum_{s_j\in Neighbor(s_i)}\lambda_{s_j} \frac{M}{\mu} \Psi_r},
\end{split}
\end{equation}
where $e_{\rm ini}(s_i)$ is the initial energy of $s_i$. As in \cite{24,53,61,78}, we define the network lifetime as the minimum lifetime of any sensor node. This is due to the fact that the failure of a single sensor node may interrupt network traffic and disable the entire UWSN. Thereby, the network lifetime can be estimated as
\begin{equation}\label{40}
\begin{split}
\Gamma_{\rm net}&=(\Gamma_{s_i})_{\rm min}=\Bigg(\frac{e_{\rm ini}(s_i)}{e_{s_i}}\Bigg)_{\rm min}\\
&=\Bigg(\frac{e_{\rm ini}(s_i) T_{\rm run}}{\lambda_{s_i} \frac{M}{\mu} \Psi_t+\sum_{s_j\in Neighbor(s_i)}\lambda_{s_j} \frac{M}{\mu} \Psi_r}\Bigg)_{\rm min},\\
&s_i\in SN=\left\{ s_1, s_2, s_3, \dots, s_{\left| SN \right|}\right\}.
\end{split}
\end{equation}

\section{Performance Evaluation}
This section demonstrates the gains of using our proposed approaches via computer simulations. The simulation setup is first described. Then we compare QLFR with four other well-known routing protocols to demonstrate its superiority performance. Finally, we present the parameter analysis to examine how the performance of the proposed routing protocol QLFR is affected by parameters.

\subsection{Simulation Configurations}
We use MATLAB to simulate and evaluate the performance of our proposed routing protocol. In the simulations, we randomly deploy sensor nodes in a 3D area of dimensions  500 m $\times$ 500 m $\times$ 500 m. Each sensor node follows the random-walk mobility pattern \cite{42}. After a sensor chooses a direction randomly, it moves to the new location at a given speed $v$. We deploy multiple sinks on the surface of the network. Once these sinks are deployed, they are stationary. Moreover, five source nodes are placed at the bottom layer of the UWSN. The maximal transmission range of a sensor node is set to $R=150$ meters; the transmission speed of acoustic waves in water is set to $v_0=1500$ m/s, and the number of sinks is set to be five. In addition, the values of energy consumption for the nodes' operations of packet transmission and reception is set to be $\Psi_t=2$ W and $\Psi_r=0.5$ W, respectively. The detailed parameters used in the simulation are shown in Table II.

In addition, the following four quantitative metrics are used to evaluate the performance of the proposed routing protocols: average~end-to-end~delay, packet delivery ratio, total energy consumption and network lifetime.

\subsection{Performance Comparison}
\subsubsection{Benchmark protocols}

In this paper, our proposed protocol QLFR is compared to four other well-known routing protocols: DBR \cite{42}, EEDBR \cite{43}, DVOR \cite{46} and QELAR \cite{53}, which represent two different paradigms of underwater routing protocols. QELAR shows how to route intelligently by using a reinforcement learning technique to balance the workload of sensor nodes. DBR, EEDBR and DVOR represent the localization-free anypath routing protocols designed without the learning process. In the following, we briefly describe these four routing protocols, and highlight some of their properties.

\begin{table}[t]\label{Table 2}
\centering
\normalsize
\caption{Values of parameters in the simulations}
 \scalebox{0.9}{
 \begin{tabular}{lll}
  \toprule
  Parameter & Description & Value \\
  \midrule
  $N$ & Number of sensor nodes in &\\
  & the entire underwater sensor&100$\sim$500\\
  & network    &\\

  $N_{\rm source}$ &Number of source nodes in the& 5 \\
  & underwater sensor network & \\

  $N_{\rm sink}$   & Number of sink nodes in the& 5\\
  & underwater sensor network&\\

  $v_0$ &Speed of acoustic wave in water& 1500 m/s\\

  $\Psi_t$ &Power of packet transmission           & 2 W \\

  $\Psi_r$              &     Power of packet reception                 & 0.5 W\\

  $R$              &     Maximal transmission range of & 150 m\\
  &a sensor node      &          \\
  $\gamma$        &     Discount factor of the long-     & \\
  &term reward for calculating   & 0.8\\
  &the Q-value & \\
  $k$              &  The difference of holding time &\\
  &between two adjacent nodes & 0.01 s$\sim$0.1 s\\
  &in the priority list & \\
  $v$             &   Movement speed of sensor nodes&\\
  &in the underwater sensor &1 m/s$\sim$5 m/s\\
  &network & \\
  \bottomrule
 \end{tabular}}
\end{table}

DBR is the first underwater sensor network routing protocol that uses node depth information to forward data packets. The essential idea behind DBR is to route data packets greedily towards the water surface in terms of depth. In addition, it uses a holding mechanism to help the forwarding candidates cooperate  with each other to select the closest forwarder. With the greedy strategy, DBR can reduce the average end-to-end delay of packet transmission.

EEDBR is an energy efficient depth-based routing protocol. In EEDBR, the next hop node is selected by first considering the residual energy of the sensor nodes. Based on residual energy, every node calculates the holding time to schedule packet forwarding. Therefore, EEDBR is an energy balanced algorithm in terms of balancing the energy consumption among sensor nodes.

DVOR is a distance-vector-based routing protocol, which uses the hop counts of sensor node towards the destination to decide on the shortest routing path. It uses a query mechanism to set up distance vectors for all nodes. The distance vectors store the least hop counts to the sink, and then data packets can be routed via the shortest path in terms of hop counts. Based on the distance vectors, DVOR can reduce detours during packet transmissions, decreasing the energy consumption and average end-to-end delay.

QELAR is a single-path routing protocol based on the Q-learning technique with the objective of  maximizing network lifetime for UWSNs. Its reward function takes into account the residual energy of each node and the energy distribution among neighboring nodes. In QELAR, the routing path are chosen for balancing the workload among sensor nodes and maximizing network lifetime. Moreover, to improve the reliability of data transmission, a retransmission mechanism after transmission failures is used in QELAR.
\\

\subsubsection{Numerical results and discussions}

\begin{figure*}[t]
\centering
\begin{minipage}[t]{0.48\textwidth}
\centering{\includegraphics[width = 1\columnwidth]{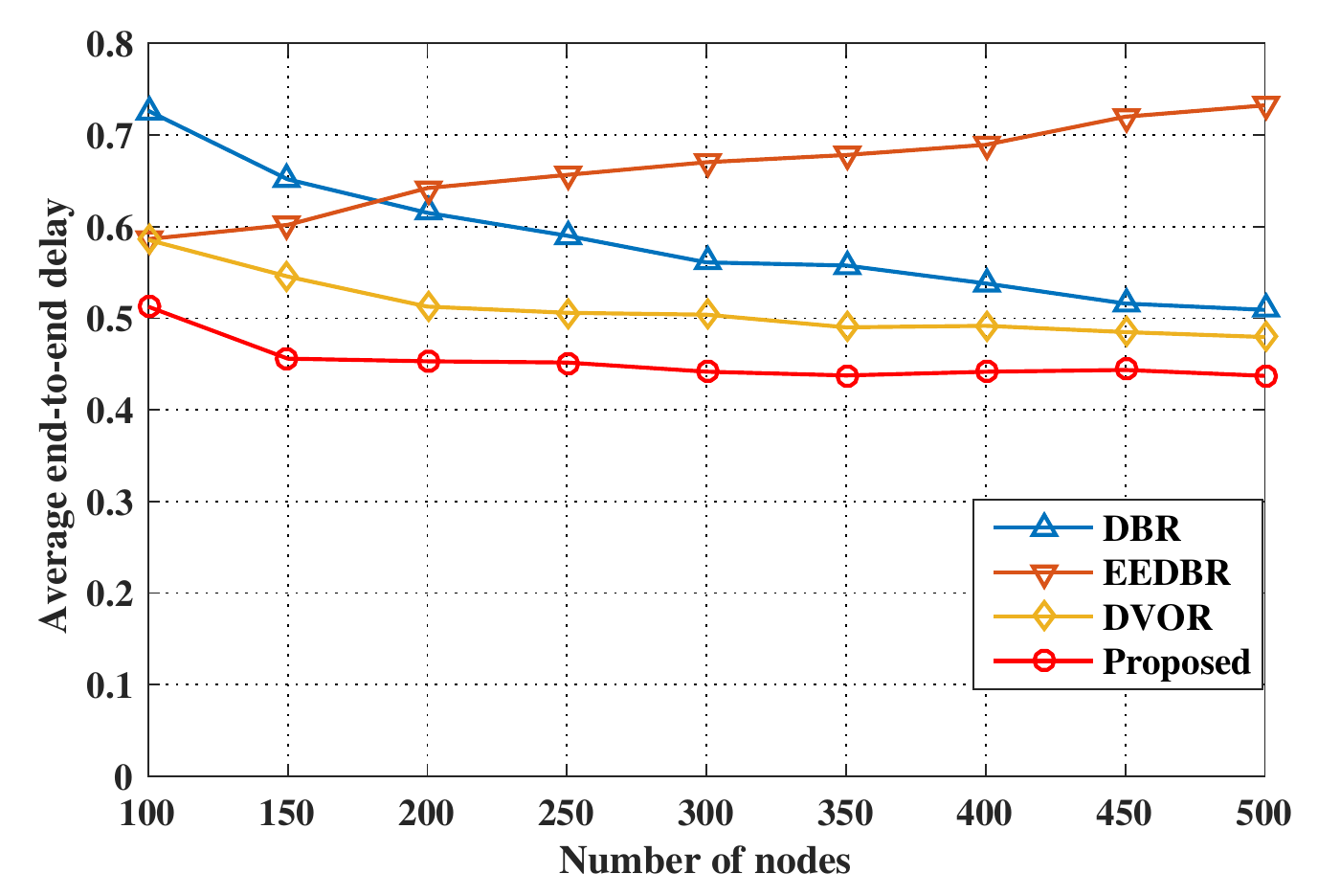}}
\caption{ Performance comparison among our proposed routing protocol, DBR, EEDBR and DVOR in terms of average end-to-end delay.}
\label{fig6}
\end{minipage}
\hspace{0.15cm}
\begin{minipage}[t]{0.48\linewidth}
\centering{\includegraphics[width = 1\columnwidth]{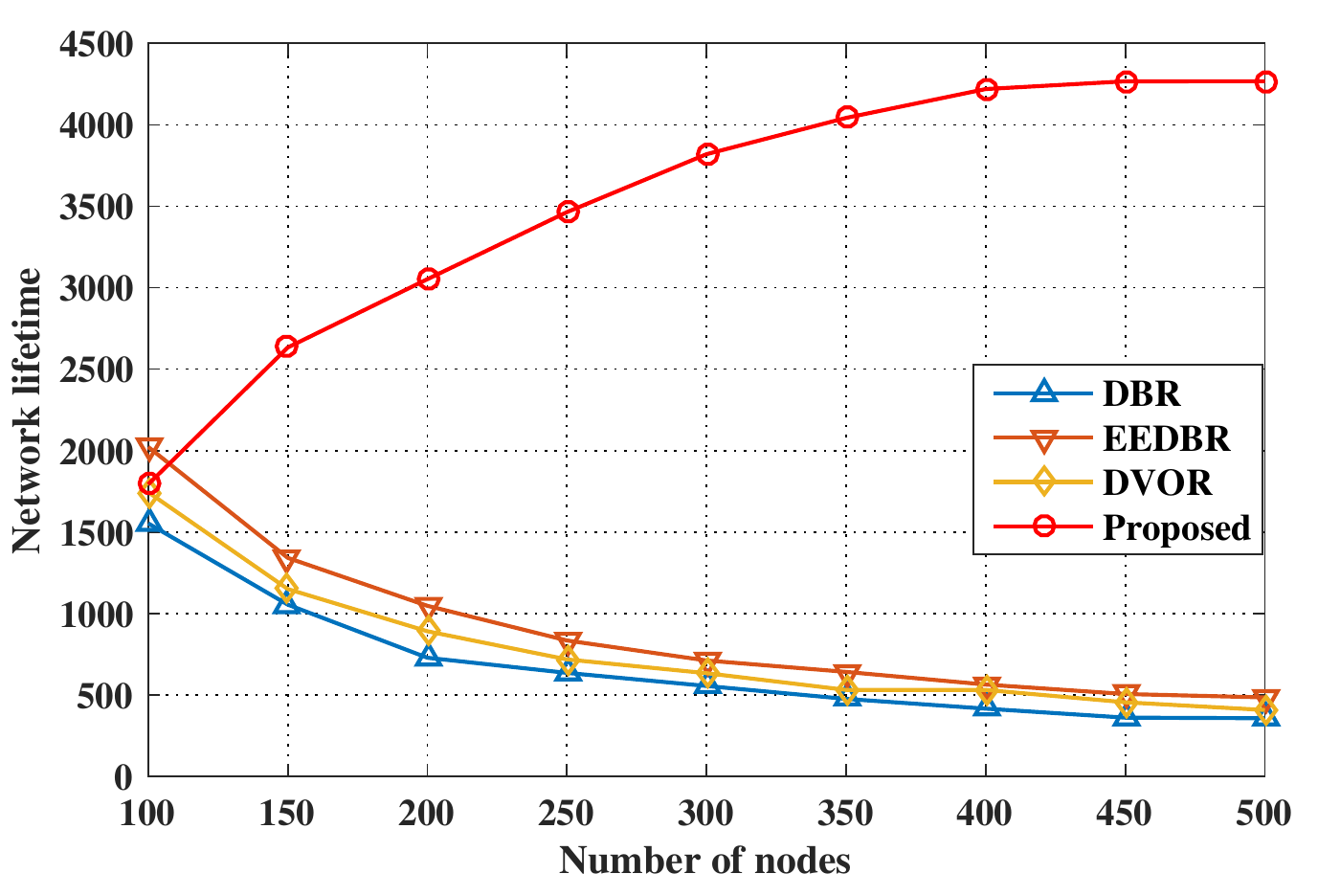}}
\caption{ Performance comparison among our proposed routing protocol, DBR, EEDBR and DVOR in terms of network lifetime. }
\label{fig7}
\end{minipage}
\end{figure*}

Firstly, we compare QLFR with DBR \cite{42}, EEDBR \cite{43} and  DVOR \cite{46} with respect of three metrics, namely the average end-to-end delay, network lifetime and packet delivery ratio. In the simulations, $k$ is set to be 0.05 s, the movement speed of a node is set to be 3 m/s, and the number of sensor nodes is set from 100 to 500.

The end-to-end delay of four schemes is shown in Fig.~\ref{fig6}. For DBR and DVOR, each sensor node will hold a packet for a certain time, which increases latency. Besides, DBR, EEDBR and DVOR only choose the forwarding candidate nodes with one-hop routing information. By contrast, QLFR benefits from the proposed Q-learning-based algorithm, which can make the global optimal routing decisions for the whole routing path. Therefore, QLFR outperforms DBR, EEDBR and DVOR in the average end-to-end delay. Moreover, as can be observed from Fig.~\ref{fig6}, the end-to-end delay decreases as network density increases in DBR, DVOR and QLFR, while it increases with the network density in EEDBR. This is owing to the fact that EEDBR uses the residual energy of the sensor nodes as a metric for scheduling packet forwarding, and thus packets need to detour in order to avoid nodes which are closer to the sink but have relatively lower energy. Therefore, for EEDBR, the more nodes in the network, the more detours will occur, which leads to an increase in end-to-end delay.

\begin{figure}[t]
\centerline{\includegraphics[width = 1\columnwidth]{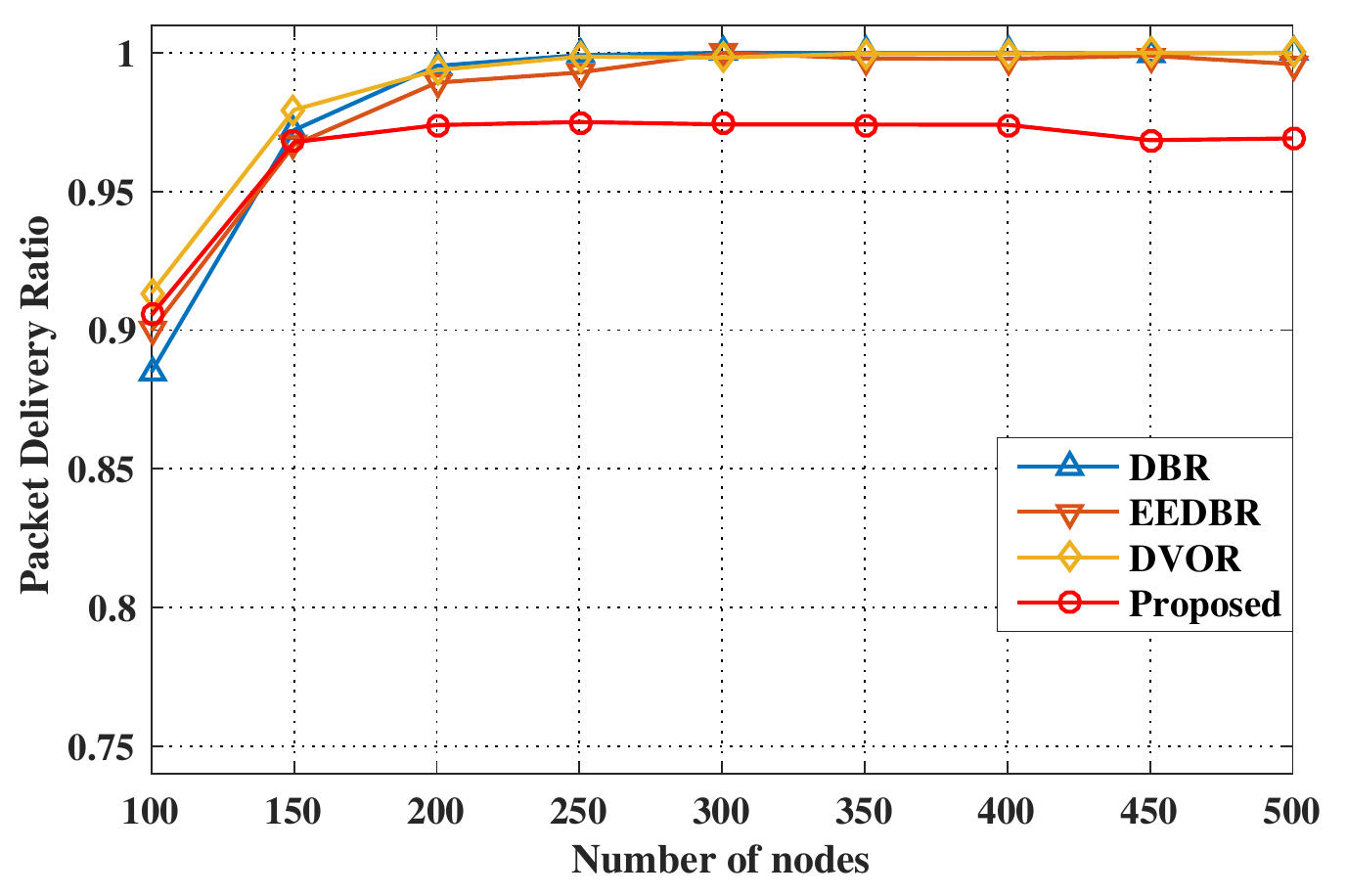}}
\caption{Performance comparison among our proposed routing protocol, DBR, EEDBR and DVOR in terms of packet delivery ratio.}
\label{fig8}
\end{figure}

The network lifetime of QLFR, DBR, EEDBR and DVOR are shown in Fig.~\ref{fig7}, which demonstrates that QLFR can always achieve the maximal lifetime compared with DBR, EEDBR and DVOR. The network lifetime presents a general trend of negative correlation with network density in EEDBR, DBR and DVOR. While it shows an opposite trend in QLFR. Here we analysis why the case is. For DBR, with the increase of the network density, a mass of redundant transmission causes excessive energy consumption, which will reduce the network lifetime. DVOR favors the shortest path to route data packets, which can reduce the energy
consumption of sensor nodes to some extent. However, the sensor nodes who lie in the shortest path are over-burdened and become hot spots due to transmitting too many packets. These hot spots fail prematurely due to energy exhaustion, shortening the lifetime of the entire UWSN. EEDBR takes the residual energy into consideration and can prolong the network lifetime compared with DBR and DVOR. However, with the increase of the network density, EEDBR will dramatically increase redundant packets due to the excessive detour. By contrast, QLFR limits the number of sensor nodes participating in packet forwarding based on a new multipath suppression scheme. Furthermore, with the energy-related cost considered in the reward function, QLFR is able to balance the workload by avoiding the chosen of nodes with relative less energy. Therefore, the proposed QLFR prolongs the network lifetime with the increase of network density, which is a significant advantage compared to DBR, EEDBR and DVOR.
\begin{figure}[t]
\centerline{\includegraphics[width = 0.94\columnwidth]{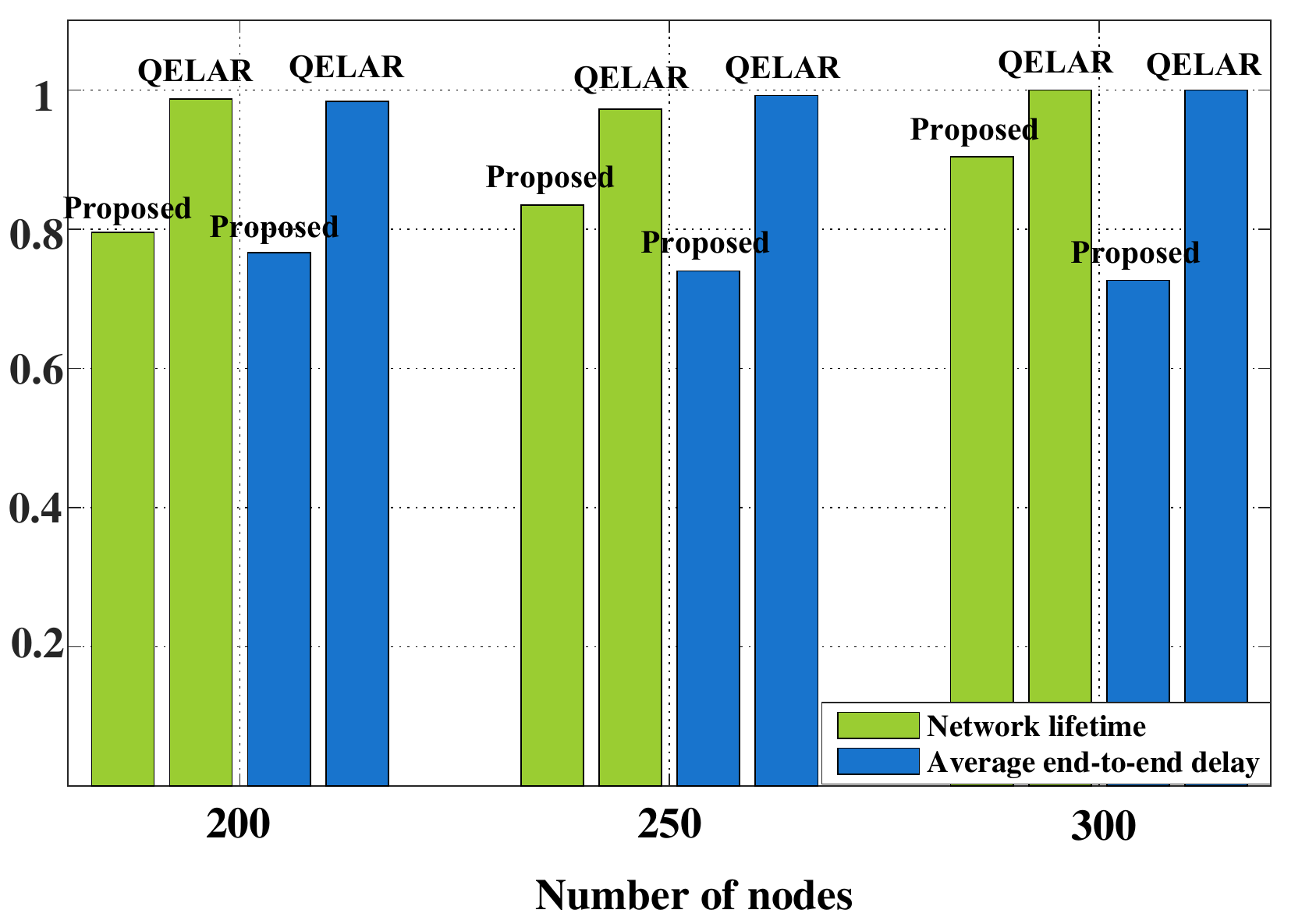}}
\caption{Performance comparison between our proposed routing protocol and QELAR.}
\label{fig9}
\end{figure}

\begin{figure*}[t]
\centering
\begin{minipage}[t]{0.48\textwidth}
\centering{\includegraphics[width = 1\columnwidth]{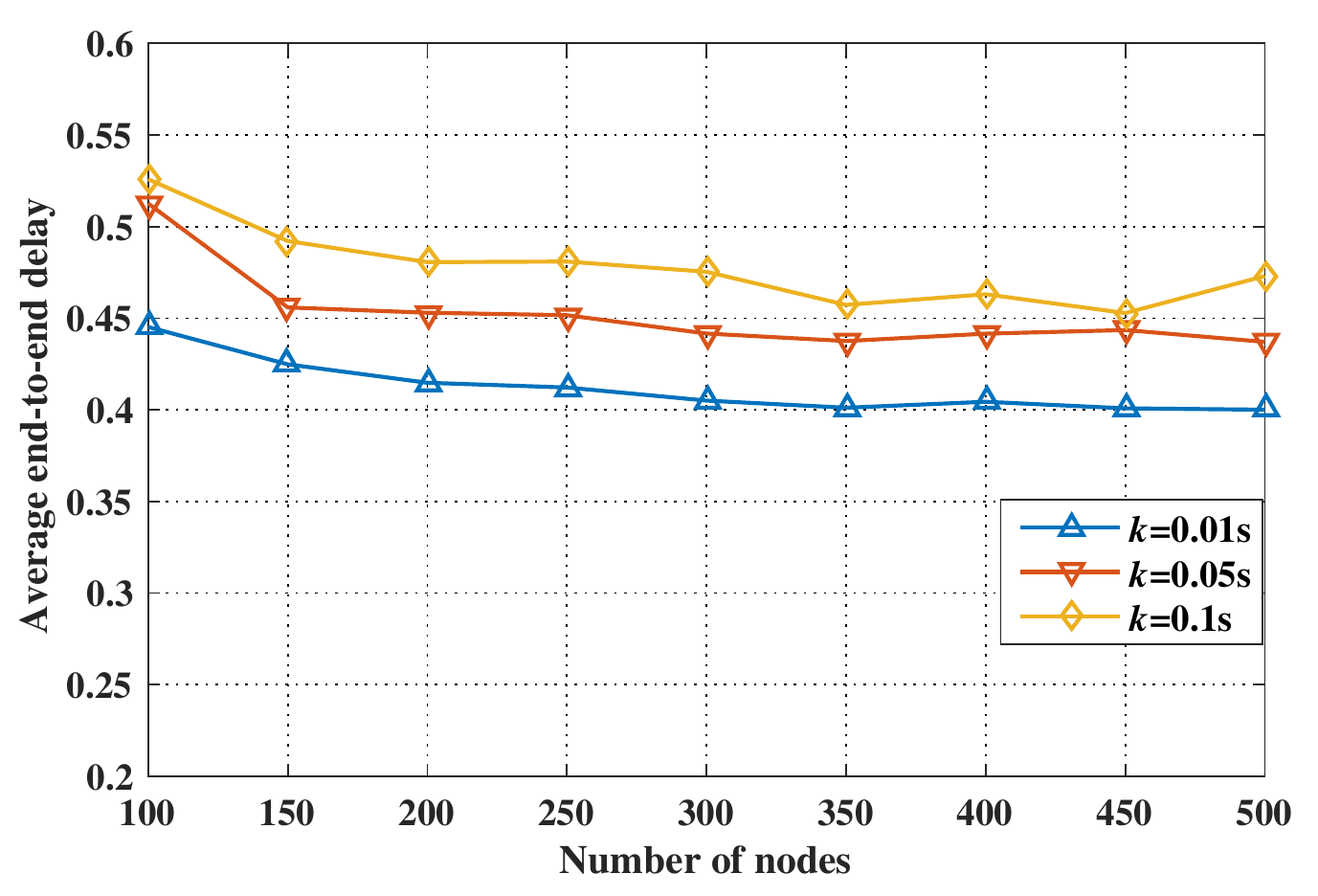}}
\caption{ Average end-to-end delay with varying $k$.}
\label{fig10}
\end{minipage}
\hspace{0.2cm}
\begin{minipage}[t]{0.48\linewidth}
\centering{\includegraphics[width = 1\columnwidth]{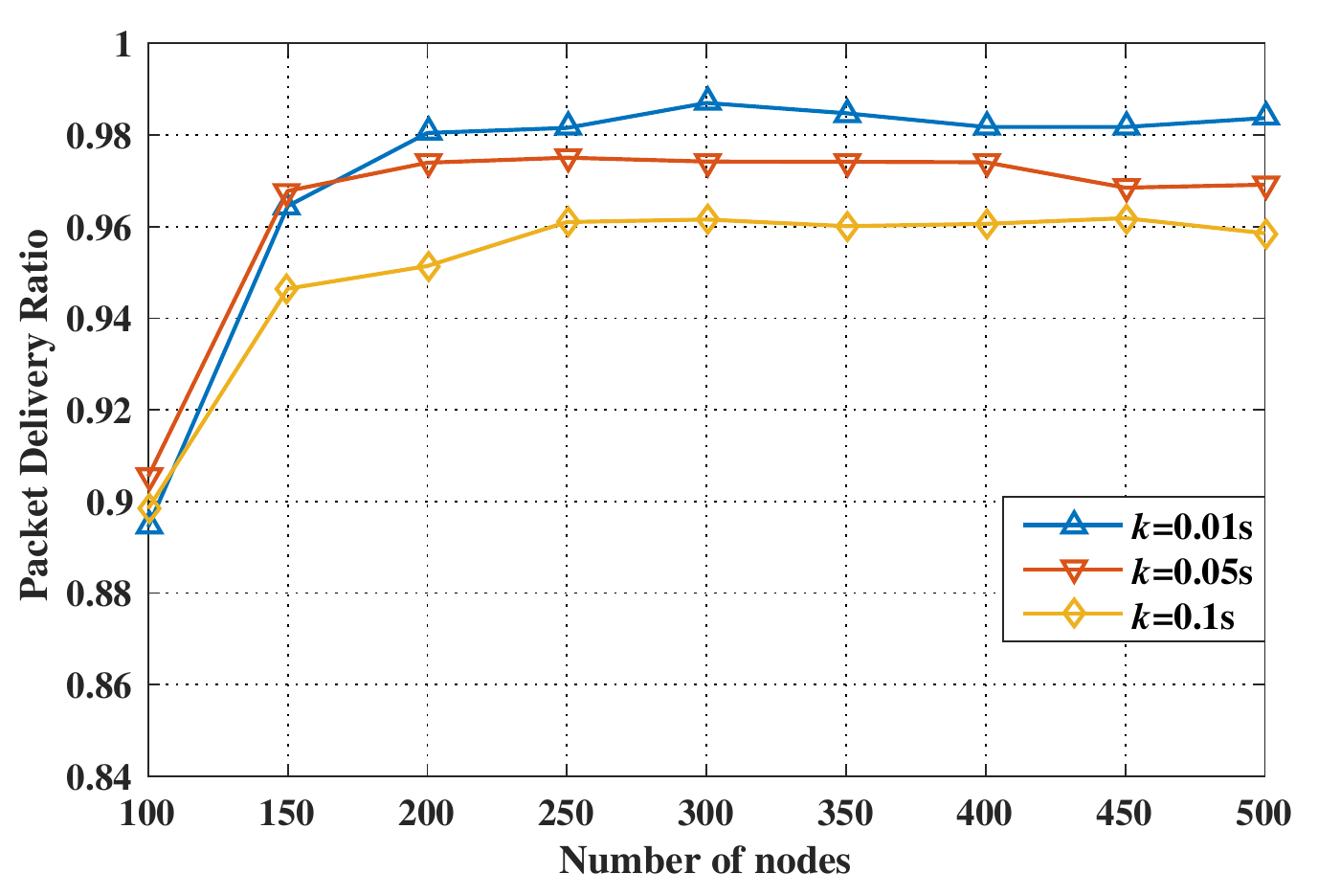}}
\caption{Packet delivery ratio with varying $k$.}
\label{fig11}
\end{minipage}
\end{figure*}
\begin{figure*}[t]
\centering
\begin{minipage}[t]{0.48\textwidth}
\centering{\includegraphics[width = 1\columnwidth]{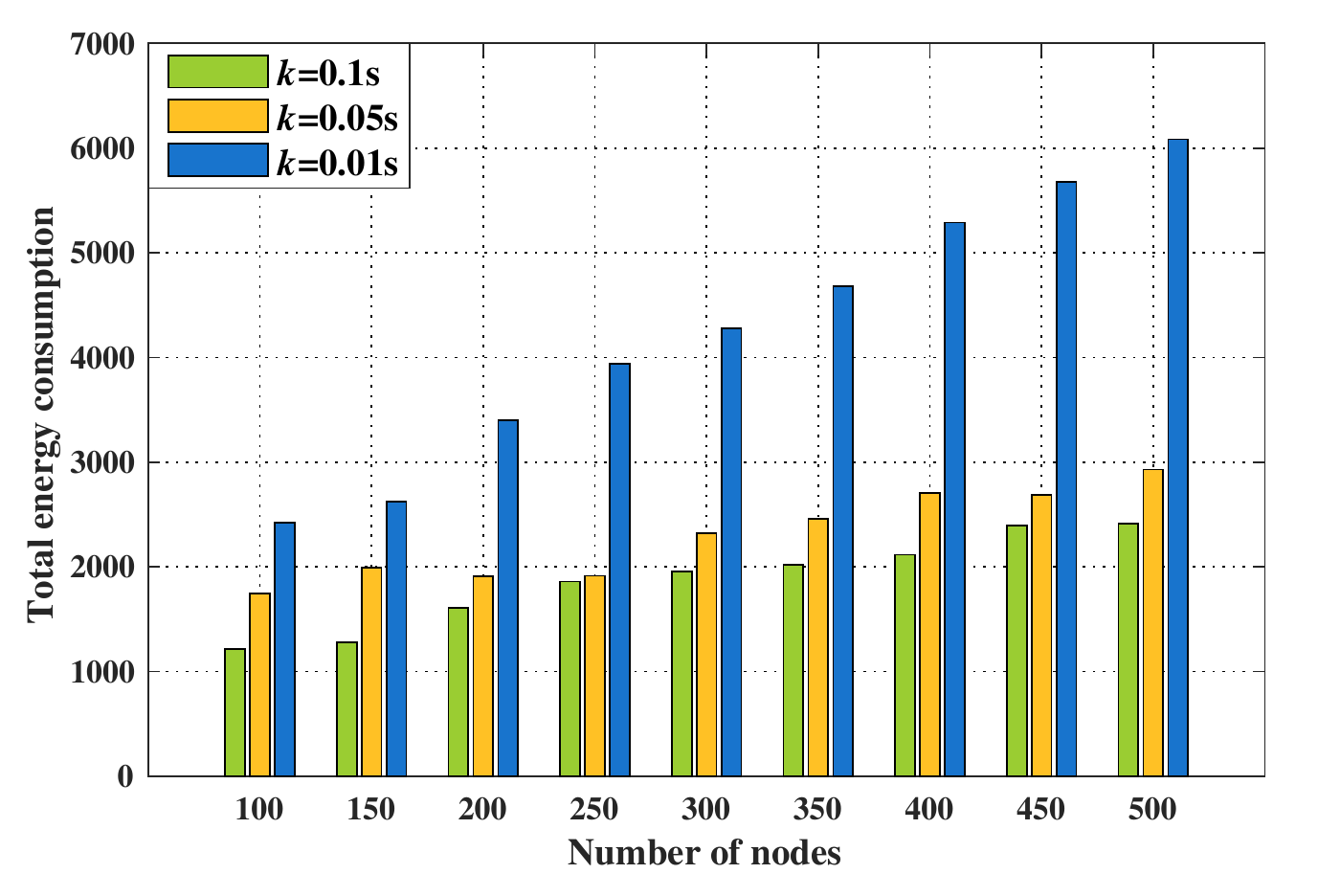}}
\caption{ Total energy consumption with varying $k$.}
\label{fig12}
\end{minipage}
\hspace{0.2cm}
\begin{minipage}[t]{0.48\linewidth}
\centering{\includegraphics[width = 1\columnwidth]{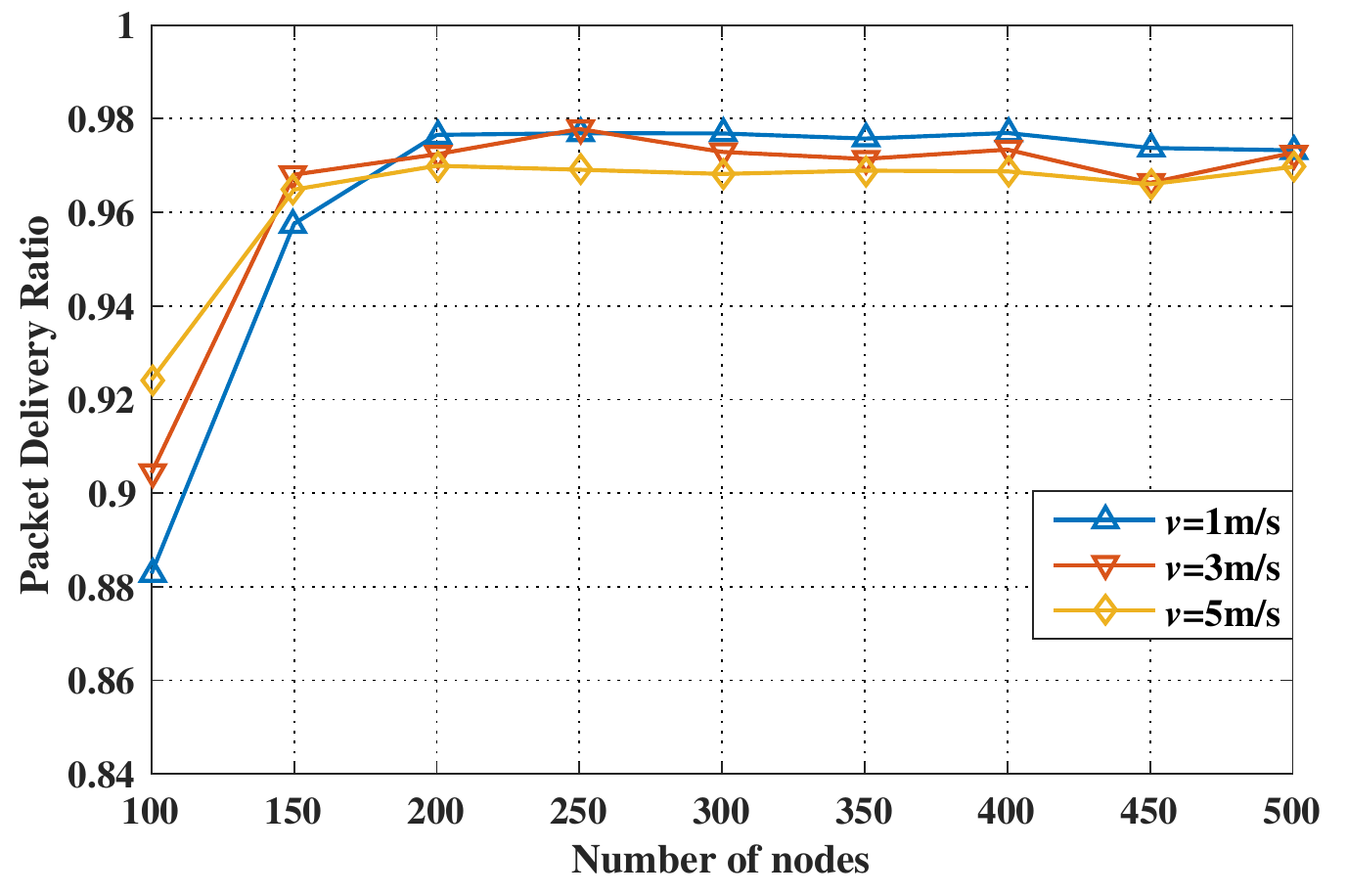}}
\caption{Packet delivery ratio with varying $v$. }
\label{fig13}
\end{minipage}
\end{figure*}

Fig.~\ref{fig8} examines the packet delivery ratio of the four methods. The results show that packet delivery ratio of EEDBR, DBR and DVOR are 2-3\% higher than that of QLFR. However, the slight improvement of the packet delivery ratio is at the expense of the network lifetime and the end-to-end delay.

To summarize, the above experimental results demonstrate that QLFR can significantly reduce the end-to-end delay and dramatically extend the network lifetime, while ensuring a packet delivery ratio almost similar to those of DBR, EEDBR and DVOR.

Now, we compare our algorithm with the Q-learning-based lifetime extended protocol, QELAR \cite{53}. The number of sensor nodes is set from 200 to 300, and the movement speed of sensor node is 1 m/s as given in \cite{53}. We evaluate the average end-to-end delay and network lifetime with different network densities.

Fig.~\ref{fig9} shows the results concerning two metrics above. To facilitate comparison, these two metrics are all normalized within 1 according to \cite{53}. As can be observed from the figure, our algorithm can significantly reduce latency compared to QELAR. The reason is that
QELAR pursues the balance of residual energy distribution as much as possible, but does not restrict end-to-end delay. Therefore, QELAR detours in the most routing paths, which will cause a huge end-to-end delay, especially in
dense networks. Furthermore, the retransmission mechanism in QELAR also increases the end-to-end delay. In contrast, by using a depth-related cost function, QLFR can reduce the latency effectively. In addition, the holding
time mechanism in QLFR can avoid retransmissions, which can further improve the end-to-end delay compared with QELAR.

As can be observed from Fig.~\ref{fig9}, the network lifetime of QLFR is slightly shorter than that of QEALR. This is because QELAR can distribute workload to almost every node in UWSNs, prolonging the network lifetime greatly. While QLFR also considers reducing the transmission delay in the network. However, the network lifetime of QLFR can usually reach more than 85\% of that in QELAR. Even on the worst case, it can still achieve 80.5\% compared with the network lifetime of QELAR.

\begin{figure*}[t]
\centering
\begin{minipage}[t]{0.48\textwidth}
\centering{\includegraphics[width = 1\columnwidth]{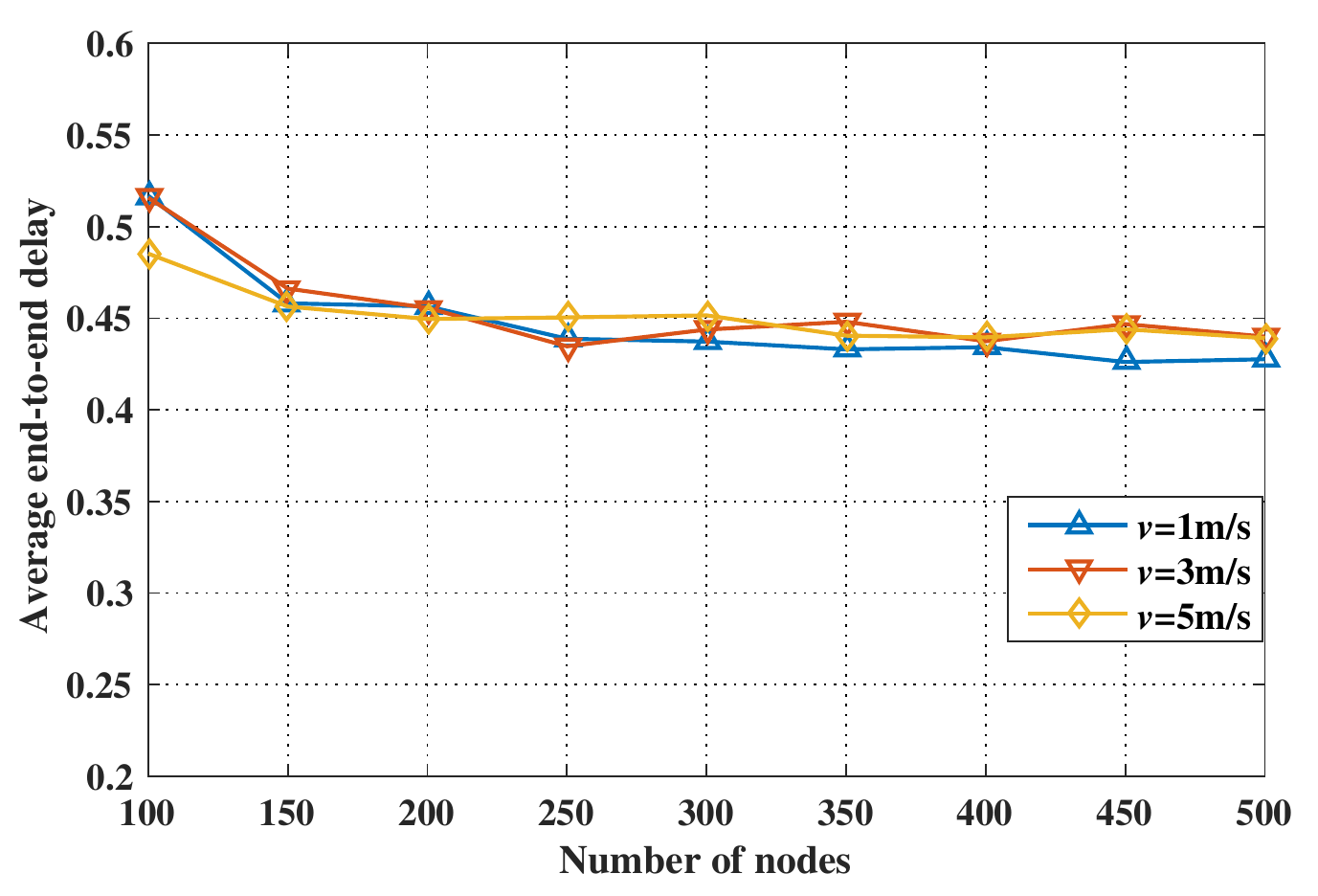}}
\caption{ Average end-to-end delay with varying $v$.}
\label{fig14}
\end{minipage}
\hspace{0.2cm}
\begin{minipage}[t]{0.48\linewidth}
\centering{\includegraphics[width = 1\columnwidth]{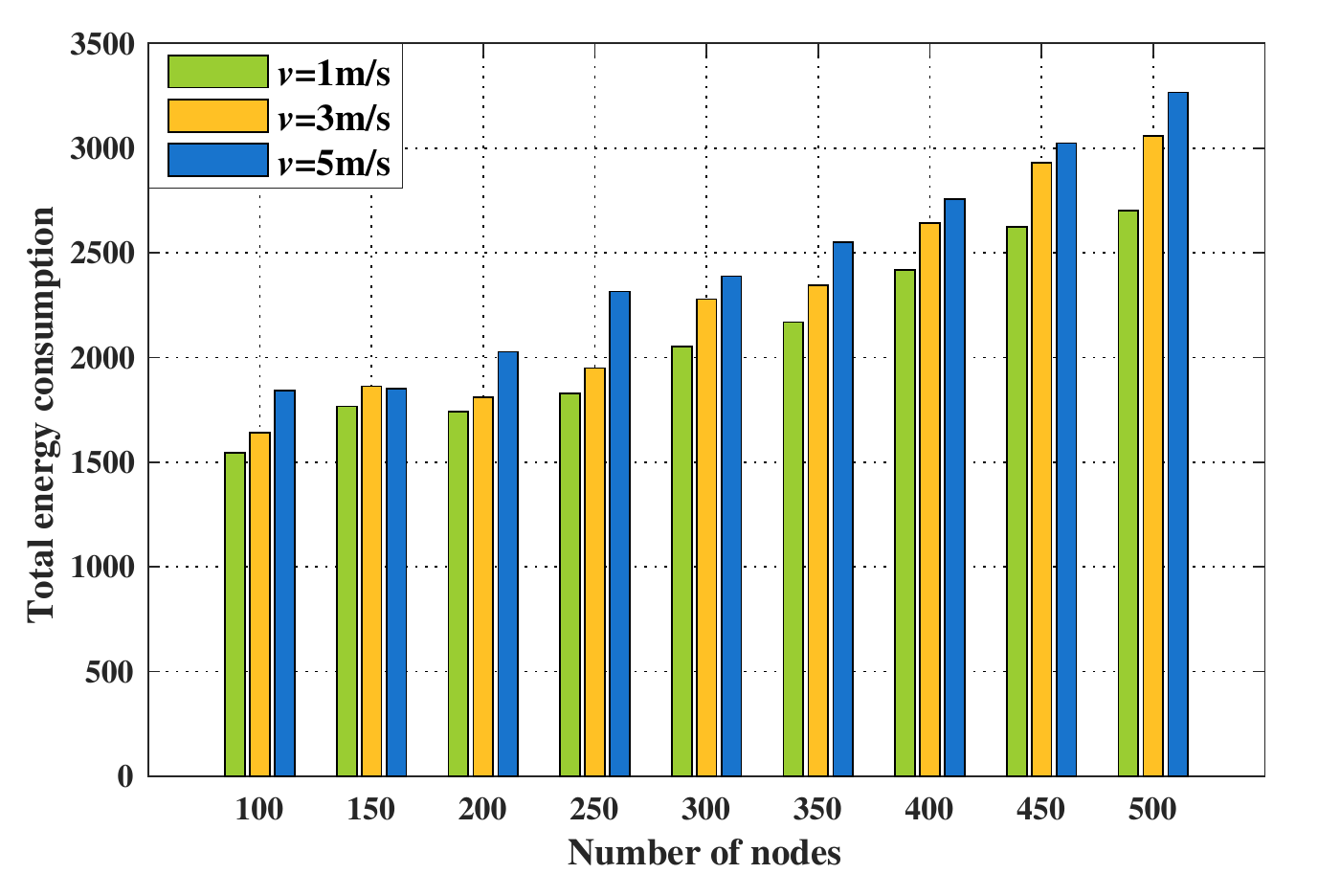}}
\caption{Total energy consumption with varying $v$.}
\label{fig15}
\end{minipage}
\end{figure*}

\subsection{Impact of Parameters}
\subsubsection{Holding time related parameter $k$}

First, we examine how the performance of QLFR is affected by $k$, which is a parameter related to the holding time as described in (9) and (15). To analyze the impact of $k$, we evaluate our algorithm in different values of $k$. Setting $R=150$ and $v_0=1500$ in (15), we have $k=\frac{0.2}{h}$; since $h\in\mathbb{N^*}$, we have $k\leq0.2$. In our experiment, we set the value of $k$ to 0.01 s, 0.05 s and 0.1 s, respectively. In addition, the movement speed of a node is set to be 3 m/s, and the number of sensor nodes is set from 100 to 500.

Fig.~\ref{fig10} shows that the average end-to-end delay with different $k$. As depicted in Fig.~\ref{fig10}, the average end-to-end delay is positively correlated to $k$. Note that the larger holding time will make sensor nodes hold the packet longer. According to (9) and (15), with the increase of $k$, the holding time of a packet will increase, and so does the end-to-end delay.

As plotted in Fig.~\ref{fig11}, the packet delivery ratio increases while $k$ decreases. This is because decreasing $k$ leads to a reduction in the holding time of nodes. Then more nodes will take part in packets forwarding, and the packet delivery ratio will increase.

As revealed in Fig.~\ref{fig12}, the total energy consumption increases with $k$ decreases. When $k=0.01$ s, the total energy consumption is much more than the other two cases. The reason is that, with a smaller $k$, nodes have shorter holding times, and hence results in less suppression of redundant packets forwarding, which leads to more energy consumption.

\subsubsection{Node movement speed $v$}
As underwater sensor nodes are mobile, we examine how the movement speed $v$ of nodes affect the performance of QLFR. To this end, we simulate QLFR with different fixed node speeds at 1 m/s, 3 m/s and 5 m/s, respectively. At the same time, $k$ is set to be 0.05 s, and the number of sensor nodes ranges from 100 to 500.

Fig.~\ref{fig13} presents the packet delivery ratio at different node speeds. As illustrated in Fig.~\ref{fig13}, the mobility of sensor nodes can improve the packet delivery ratio of the protocol, especially in sparse networks. This can be explained by the fact that when movement speeds of sensor nodes increase, the network topology will change rapidly. In this case, due to the rapid movement of nodes, the void-hole region in network can be covered rapidly and the coverage rate of the network will increase in the unit time. Therefore, if the network density is low, the delivery ratio increases with the increase of the node speed. However, if the network becomes dense, the probability of void-hole region emerging will be reduced, hence, the movement speed of nodes has little impact on packet delivery ratio.

As shown in Fig.~\ref{fig14}, the movement speed of nodes has little impact on the average end-to-end delay. As can be seen from Fig.~\ref{fig15}, the total energy consumption only increases slightly with  the movement speed of nodes. This is because the node movement can increase the coverage of the network and then slightly raise the data packet transmission rate in the network. However, the change of total energy consumption at different  movement speeds of nodes is gentle and negligible.

Based on the above analysis, QLFR can well deal with the mobility of sensor nodes and is suitable for mobile underwater sensor networks which feature highly dynamic network topology.

\section{Conclusion }
This paper investigated routing protocols in underwater sensor networks (UWSNs) and proposed a Q-learning-based localization-free anypath routing protocol dubbed QLFR. First, in order to extend the network lifetime and to reduce the end-to-end delay, we designed the energy-related and depth-based rewards to calculate the Q-value, which is considered as the priority metric for forwarding candidate nodes. Then a new holding time mechanism was proposed to schedule packets transmission operation among candidates according to the priority levels of them. At last, we proposed a new scheme to control the length of the priority list, so as to further reduce unnecessary transmissions. Extensive simulation results were presented to demonstrate that our routing protocol outperforms the comparative routing protocols.
\bibliographystyle{IEEEtran}
\bibliography{reference}

\end{document}